\documentclass[12pt,a4paper]{article}
\usepackage{bbm}
\usepackage{amsmath}
\usepackage{amssymb}
\usepackage{graphicx}
\usepackage{psfrag}
\begin{document}

\begin{flushright}
\today \\
BI-TP 2001/14 \\
\end{flushright}

\begin{center}
\large{\bf Spectrum of the Dirac Operator coupled to two-dimensional
quantum gravity}
\end{center}

\begin{center}
L. Bogacz$^{1,2}$, Z. Burda$^{1,2}$, C. Petersen$^{1}$ and
B. Petersson$^{1}$
\end{center}

\centerline{$^{1}$Fakult\"at f\"ur Physik, Universit\"at Bielefeld}
\centerline{P.O.Box 100131, D-33501 Bielefeld, Germany}
\vspace{0.3cm}
\centerline{$^{2}$Institute of Physics, Jagellonian University}
\centerline{ul. Reymonta 4, 30-059 Krakow, Poland}

\begin{abstract} \normalsize \noindent

We implement fermions on dynamical random triangulation
and determine numerically the spectrum of the Dirac-Wilson
operator ${\cal D}$ for the system of Majorana fermions
coupled to two-dimensional Euclidean quantum gravity.
We study the dependence of the spectrum
of the operator $\epsilon {\cal D}$ on the
hopping parameter. We find that the distributions of 
the lowest eigenvalues become discrete when the hopping 
parameter approaches the value $1/\sqrt{3}$. We show that
this phenomenon is related to the behavior of the system
in the 'antiferromagnetic' phase of the corresponding 
Ising model.
Using finite size analysis we determine critical exponents
controlling the scaling of the lowest eigenvalue
of the spectrum including
the Hausdorff dimension $d_H$ and
the exponent $\kappa$ which tells us how fast
the pseudo-critical value of the hopping parameter approaches
its infinite volume limit.

\end{abstract}

\section*{Introduction}
The dynamical triangulation approach to quantum gravity
has proven to be a very powerful method \cite{d1,kkm,d2}.
In two-dimensions it yields the same results for
critical exponents as the Liouville theory \cite{kpz,d3,dk}. Contrary
to the latter, this approach can be straightforwardly generalized
to higher dimensional case which is frequently referred to as simplicial
gravity \cite{am,aj}. Results from numerical studies
of pure gravity without matter fields in four dimensions
showed that the continuum limit of this model does not exist
\cite{bbkp}. In order to obtain more realistic models, one has tried
to include matter fields and to couple them to gravity \cite{bbkptt}.
This program has so far succeeded only for bosonic matter.
Putting fermions on random simplicial manifold
is a more difficult task. In general it requires introducing
an additional field of local frames in order to define a spin structure
\cite{r,bm,bjk}.
In the case of a compact manifold this is a topological problem.
Although many ingredients of the construction
are known and can be generalized to any number of dimensions,
the topological part of the problem
has been solved so far only in two dimensions \cite{bjk,bbjkpp}.

In this paper we will study properties of the Dirac-Wilson
operator on two-dimensional dynamical triangulation with
spherical topology. The analysis of the spectrum in the
critical region allows us to calculate critical indices
as for example the Hausdorff dimension. 

We cross-check properties of the spectrum using the fact that 
the partition function of the
fermionic model can be mapped into the 
partition function of Ising model on dynamical 
triangulation, which is analytically solvable \cite{k,bk,bj}.

The spectrum of the Majorana-Dirac-Wilson operator 
$\epsilon {\cal D}$ becomes discrete when the hopping 
parameter admits the value $1/\sqrt{3}$ corresponding 
to the value $\beta=0$ of the coupling in the Ising model. 
We show that this behaviour can be explained by the presence
of a set of points in 'antiferromagnetic' phase ($\beta<0$),
for which some eigenvalues
of the operator are determined by local properties 
of the triangulation and not by its random character.

The paper is organized as follows~: First
we define the model, then we recall some facts
about its relation to the Ising model \cite{bm}, we present
results of numerical studies and shortly
conclude at the end by summarizing and listing
open questions. In the appendix, for comparison,
we calculate the spectrum of the Dirac-Wilson operator
on a regular triangulation.

\section*{The model}
The model of fermions minimally coupled to Euclidean gravity
is given by the partition function
\begin{equation}
{\cal Z} = \sum_{T \in {\cal T}} {\cal Z}_T
= \sum_{T \in {\cal T}} \int
\prod_i d\bar{\Psi}_i d\Psi_i \ e^{-S_T}
\label{Z}
\end{equation}
where the sum goes over $d$-dimensional simplicial
manifolds from a class ${\cal T}$, say, for instance, with
spherical topology. Each triangulation
is dressed with the fermion field 
located in the centers of $d$-simplices.
The integral over field on a given triangulation $T$
defines the partition function ${\cal Z}_T$, which
at the same time provides a weight of this
triangulation in the ensemble. The action reads
\begin{equation}
S_T = -K \sum_{\langle ij \rangle} 
\bar{\Psi}_i {\cal H}_{ij} \Psi_j +
\frac{1}{2} \sum_i \bar{\Psi}_i \Psi_i \, =  \,
\sum_{i,j} \bar{\Psi}_i {\cal D}_{ij} \Psi_j \, ,
\label{sh}
\end{equation}
where the fermionic fields $\Psi_i$ are located 
in the centers of triangles. The sum over  
$\langle ij \rangle$ runs over
oriented pairs of neighboring triangles,
or equivalently, over oriented dual links.
The hopping operator is 
\begin{equation}
{\cal H}_{ij} = \frac{1}{2} ( 1 + n^{(i)}_{ij}\cdot \gamma ) \,
{\cal U}_{ij} \, .
\label{H}
\end{equation}
The Dirac-Wilson operator is denoted by ${\cal D}_{ij}$
and the spin connection by ${\cal U}_{ij}$.
In order to be able to calculate spinor and vector components,
we endow each $d$-simplex with an orthonormal
local frame. A frame is a set of orthonormal oriented vectors
$e_a$, $a=1,\dots d$. To each vector $e_a$ we
ascribe a Dirac gamma matrix $\gamma^a$, in such a way that
its numerical value is identical in each frame.
The local vector $n_{ij}$ in eq. (\ref{H}) is
a unit vector which points from the center of the simplex
$j$ to the center of one of its neighbors $i$. It just
tells us the direction of the local derivative.
The inner product of this vector and of gamma matrices,
which is denoted by dot in (\ref{H}), has to be understood as a
sum of gamma matrices $\gamma^a$
multiplied by the components of $[n^{(i)}_{ij}]_a$
in the given frame at $i$, denoted
by the upper index. Thus, the
product of the same vector $n_{ij}$ expressed in another frame
yields a different matrix~: $n^{(j)}_{ij} \cdot \gamma =
[n^{(j)}_{ij}]_a \gamma^a$.

As mentioned the matrix ${\cal U}_{ij}$ plays the role
of spin connection. It allows us to parallel transport a
spinor from the simplex $j$ to the simplex $i$, or in other words,
to recalculate spinor components between
two neighboring frames $i$ and $j$. The matrix ${\cal U}_{ij}$
is an image in the spinorial representation
of the rotation matrix $U_{ij}$ which parallel
transports vectors.
The map $U_{ij} \rightarrow {\cal U}_{ij}$ is not unique, namely
it is defined up to sign. As we will see below, the
signs of ${\cal U}$ must be adjusted to fulfill
a consistency condition (\ref{sp})
for all elementary plaquettes of the simplicial manifold.
This is a topological problem.

This problem has been solved in
two-dimensions where an explicit construction
of the signs of the spin connection matrices ${\cal U}_{ij}$ has been
given \cite{bjk}. Let us shortly recall the main steps
of the construction.

In two dimensions each orthonormal frame
consists of two vectors $e_{ia}$ where $a$ is $1$ or $2$.
The first index of $e_{ia}$ refers to
the triangle in which the frame is located.
For any pair of neighboring
triangles $i$, $j$ we can define a spin connection
as a two by two rotation matrix $[U_{ij}]^a_b$,
such that\footnote{In general a connection
can be a dynamical field.}
$e_{ia} = \sum_b [U_{ij}]_a^b e_{jb}$.
Using matrix notation this relation can be written as
$e_i = U_{ij} e_j$, where
\begin{equation}
U_{ij} = e^{\epsilon \Delta \phi_{ij}} = \left( \begin{array}{rr}
\cos \Delta \phi_{ij} & \sin \Delta \phi_{ij} \\
-\sin\Delta \phi_{ij} & \cos \Delta \phi_{ij} \end{array} \right)
\label{u0}
\end{equation}
and $\Delta \phi_{ij}$ is
the relative angle between the
two neighboring frames. $\epsilon$ is the standard antisymmetric
tensor.

The trace of an elementary loop around a dual plaquette is
a geometrical invariant directly related to the curvature
(deficit angle) of the vertex in the center of the plaquette.
One can check that
\begin{equation}
\frac{1}{2} {\rm Tr} \, U U \dots U =
\frac{1}{2} {\rm Tr} \, e^{\epsilon (2\pi - \Delta_P) } =
\cos \Delta_P \, ,
\label{UUU}
\end{equation}
where $\Delta_P$ is the deficit angle of the vertex
in middle of the plaquette.
The product $U U \dots U$ of connections on all links on the
plaquette perimeter $P$ is a rotation
matrix which gives the integrated rotation
of a tangent vector parallel transported around this loop.
The equation (\ref{UUU}) is a sort of
Wilson discretization \cite{w1,w2} of curvature calculated from
the Cartan structure equations \cite{top}.

\begin{figure}
\begin{center}
\psfrag{i}{$i$}
\psfrag{j}{$j$}
\psfrag{ei}{$e_{i1}$}
\psfrag{ej}{$e_{j1}$}
\psfrag{n}{$n_{ji}$}
\psfrag{m}{$n$}
\psfrag{k}{$k$}
\psfrag{B}{$$}
\includegraphics[width=8cm]{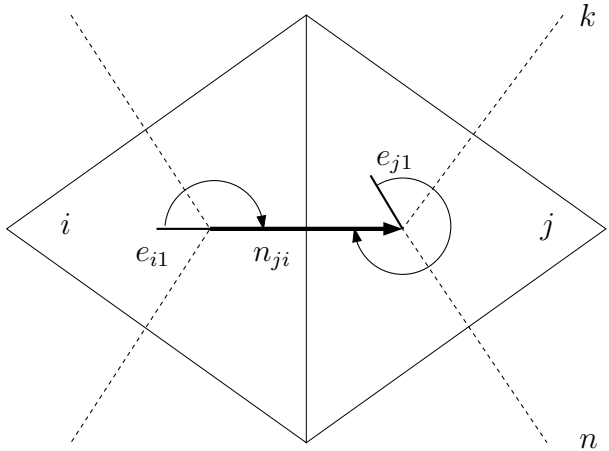}
\caption{\label{fig1}
Local geometry of two neighboring triangles is shown.
The position of the first frame vector $e_{1}$ for a given
triangle is marked by a line emerging from the triangle center.
The position of the second frame vector $e_{2}$ is
implicitly given by the fact that the angle between
$e_1$ and $e_2$ counted clockwise is $\pi/2$.
The vector $n_{ji}$ points from the center of the triangle $i$
to $j$.
The arch in the triangle $i$ represents the angle
$\phi^{(i)}_j$ between $e_{i1}$ and $n_{ji}$.
The arch in the triangle $j$ corresponds to
the angle $\phi^{(j)}_i$ between $e_{j1}$ and $n_{ij}$.
In the example shown in figure $\phi^{(i)}_j=\pi$,
$\phi^{(j)}_i=5\pi/3$, and for other two neighbors of $j$~:
$\phi^{(j)}_k=\pi/3$, $\phi^{(j)}_n=\pi$.}
\end{center}
\end{figure}

Now the idea is to write down an analogous equation
as (\ref{UUU}) in the spinorial representation. First we have to
introduce a parallel transporter ${\cal U}_{ij}$ for
spinors for each pair of neighboring vertices.
This is exactly the object which we need in (\ref{H}). The connection
${\cal U}_{ij}$ is an spinorial image of $U_{ij}$.
One can choose a representation
of gamma matrices such that $U_{ij} = {\cal U}_{ij}^2$.
One immediately sees that indeed ${\cal U}_{ij}$
can be calculated for a given $U_{ij}$ up to sign. When
defining the Dirac-Wilson operator (\ref{H}) we cannot allow
for ambiguities, so we have to give a unique prescription
how to calculate ${\cal U}_{ij}$. We do this by choosing
\begin{equation}
{\cal U}_{ij} = e^{\frac{\epsilon\Delta \phi_{ij}}{2}}
= \left( \begin{array}{rr}
\cos \frac{\Delta \phi_{ij}}{2} & \sin \frac{\Delta \phi_{ij}}{2} \\ & \\
-\sin \frac{\Delta \phi_{ij}}{2}
& \cos \frac{\Delta \phi_{ij}}{2} \end{array} \right)
\label{u1}
\end{equation}
and specifying the angles $\Delta \phi_{ij}$ uniquely.
More precisely we define
$\Delta \phi_{ij} = \phi^{(j)}_i - \phi^{(i)}_j + \pi$
where the angle $\phi^{(j)}_i$ at triangle $j$ is the angle between
the vector $e_{j1}$ of the frame at $j$ and the vector
$n_{ij}$ (pointing from $j$ to $i$), and
likewise $\phi^{(i)}_j$ at triangle $i$ is the angle between
the vector $e_{i1}$ and the vector $n_{ji}$
(pointing from $i$ to $j$) (see fig.\ref{fig1}).
Both the angles are restricted
to the range $[0,2\pi)$ and both are measured in
the same direction, say clockwise.
Thus the angle $\Delta \phi_{ij}$ is defined without
the $2\pi$ ambiguity and hence the rotation matrix ${\cal U}_{ij}$
is also uniquely determined including the total sign.

One can easily check that, for the definition (\ref{u1})
of ${\cal U}_{ij}$'s, the parallel transporter
around an elementary loop gives
\begin{equation}
\frac{1}{2} {\rm Tr} \, {\cal U} {\cal U} \dots {\cal U}
= S_P \cos \frac{\Delta_P}{2} \, .
\label{p}
\end{equation}
The argument of cosine got halved $\Delta_P \rightarrow \Delta_P/2$
in comparison with (\ref{UUU}) because for each link on $P$ we
have ${\cal U}^2 = U$. The total sign $S_P$
of the product ${\cal U}{\cal U} \dots {\cal U}$ has to be
calculated. It turns out to depend on all angles
$\Delta \phi_{ij}$ on the loop and it may admit either
value $\pm 1$ \cite{bjk}.

The presence of elementary plaquettes
which would have negative sign is an unwanted effect.
For example, a spinor transported around a flat plaquette
with $S_P=-1$ would change the sign $\psi \rightarrow -\psi$.
We require that the parallel transport around a close
loop in a flat patch not change a spinor. Furthermore,
we require the signs $S_P$ to be positive
for all elementary plaquettes
\begin{equation}
S_P = +1, \quad \forall P \, .
\label{sp}
\end{equation}
One can give the following argument in favor of the naturalness
of this requirement.
An elementary loop goes through triangles sharing a vertex. The
geometry of a patch consisting of those triangles corresponds
to the geometry of a cone. It is everywhere flat except at
the vertex where it is singular.
One can regularize such geometry by smoothing
the peak of the cone (making it differentiable)
in a very small region with radius $\epsilon \approx 0$.
Such a regularization does not affect the loop which
lies in a distance $R \gg \epsilon$
from the vertex. Continuously shrinking the loop in such
a regularized geometry, one can continuously change the
angle of the loop rotation matrix without changing the
sign. A completely shrunken loop
must have positive sign since it lies in a flat patch.
This implies $S_P = +1$. One can also check that the consistency
condition (\ref{sp}) plays an essential role
in the topological considerations or in deriving the equivalence
with the Ising model.

The construction of the connections given in (\ref{u1})
does not fulfill the consistency condition (\ref{sp}).
We will therefore modify the construction of ${\cal U}$'s by
introducing for each link an additional sign degree of
freedom $s_{ij}$
\begin{equation}
{\cal U}_{ij} = s_{ij} e^{\frac{\epsilon\Delta \phi_{ij}}{2}} \, .
\label{u}
\end{equation}
One can show that this freedom is sufficient to globally, for
each elementary loop, fulfill the consistency condition
(\ref{sp}), on a triangulation of an orientable manifold.
Thus, technically, to define the Dirac-Wilson operator
on a triangulation, we have to first assign
an orthonormal frame to each triangle, and then for
the frame assignment, to find link signs
$s_{ij}$ meeting the consistency condition $(\ref{sp})$
for each plaquette. The remaining part is straightforward.
Namely, we express the operator ${\cal U}_{ij}$ in terms
of the angles $\phi^{(j)}_i$ and $\phi^{(i)}_j$ and likewise,
the product $n^{(i)}_{ij} \cdot \gamma$ in terms of
$\phi^{(i)}_j$. Thus, we parameterize
the hopping operator ${\cal H}_{ij}$
entirely by $\phi^{(j)}_i$ and $\phi^{(i)}_j$ and $s_{ij}$.
For each pair of neighboring triangles the angles
can be read off from the given
frame assignment (see fig.\ref{fig1}).

Choosing the following representation of gamma matrices~:
$\gamma_1 = \sigma_3$, $\gamma_2=\sigma_1$, where the $\sigma$
are Pauli matrices, we eventually arrive at
\begin{equation}
{\cal H}_{ij} = s_{ij} \cdot
\left(\begin{array}{rr}
 \sin \frac{\phi^{(i)}_j}{2}  \cos \frac{\phi^{(j)}_i}{2} &
 \sin \frac{\phi^{(i)}_j}{2}  \sin \frac{\phi^{(j)}_i}{2} \\ & \\
-\cos \frac{\phi^{(i)}_j}{2}  \cos \frac{\phi^{(j)}_i}{2} &
-\cos \frac{\phi^{(i)}_j}{2}  \sin \frac{\phi^{(j)}_i}{2}
\end{array}\right) \, .
\label{Hm}
\end{equation}
We see that in two dimensions the Dirac-Wilson operator
on a triangulation $T$ is given by
a matrix consisting of two-by-two blocks
\begin{equation}
\left[{\cal D}_{ij}\right]_{\alpha}^{\beta} = 
\frac{1}{2} \delta_{ij} \delta_{\alpha}^{\beta} -
K P_{ij} \left[{\cal H}_{ij}\right]_\alpha^\beta 
\label{dwo}
\end{equation}
where $P_{ij}$ is an adjacency matrix~:
\begin{equation}
P_{ij} = \left\{\begin{array}{ll} 1, & {\rm if \quad} i,j 
{\rm \quad are \ neighbors} \\
0, & {\rm otherwise} \end{array} \right.
\end{equation}
There is no summation over $ij$ in the equation
(\ref{dwo}). The blocks ${\cal H}_{ij}$ have a very simple
structure. In fact, we can simplify it further by restricting
the set of values of the angles in (\ref{Hm})
from the whole interval $[0,2\pi)$ to a discrete
set of three values separated by $2\pi/3$, for instance,
$\pi/3,\pi,5\pi/3$. For this choice,
the first vector $e_{1}$ of a frame at a triangle
points from the center of the triangle to one of its
vertices. In a sense, this set of three frame positions
is a minimal set reflecting the symmetry of equilateral triangle.

Since physical quantities do not depend on the choice of frames,
this restriction is a sort of gauge condition. With this
choice, the blocks (\ref{Hm}) may admit only nine different
forms depending on the nine different choices of the
angles $\phi_i^{j}$, $\phi_j^{i}$ in (\ref{Hm}). 
They can be precomputed.
For example, for the frame assignment in fig.\ref{fig1}
$\phi^{(j)}_i = {5\pi}/3$, $\phi^{(i)}_j = \pi$ and for
$s_{ij}=1$ we have
\begin{equation}
{\cal H}_{ij} = \left(\begin{array}{rr} -\frac{\sqrt{3}}{2}  &  \frac{1}{2} \\
                      0 &  0\end{array}\right) \quad , \quad
{\cal H}_{ji} = \left(\begin{array}{rr} 0  &  -\frac{1}{2} \\
                      0 & -\frac{\sqrt{3}}{2} \end{array}\right) \, .
\end{equation}

\section*{Fermions and the Ising model}

The idea is now to calculate
spectra of the Dirac-Wilson
operator for different triangulations
from the ensemble (\ref{Z}).
Summing up (averaging) all the spectra
we obtain the spectrum
of the Dirac-Wilson operator for fermions interacting
with 2d gravity. More precisely, we will consider a field
of Majorana-fermions coupled to gravity.
At the critical point it corresponds
to the conformal field with the central charge $c=1/2$.

Denote the components of the spinor $\Psi$ by $\Psi_\alpha$,
and of $\bar{\Psi}$ by $\Psi^\beta$.
The Majorana condition reads~:
$\Psi^\beta = \epsilon^{\beta\alpha} \Psi_\alpha$ or
$\Psi_\alpha = \Psi^\beta \epsilon_{\beta\alpha}$,
where $\epsilon$ is the standard antisymmetric tensor,
In this notation, the action for Majorana fermions
can be written as
\begin{equation}
S_T = \sum_{ij} \Psi_i^\alpha [{\cal D}_{ij}]_\alpha^\beta \Psi_{i\beta} =
\sum_{ij} \Psi_{i\alpha} \widehat{\cal D}_{ij}^{\alpha\beta} \Psi_{j\beta}
\, ,
\end{equation}
where
\begin{equation}
\widehat{\cal D}_{ij}^{\alpha\beta} = \,
\epsilon^{\alpha \gamma} [{\cal D}_{ij}]_\gamma^\beta \, = \,
\frac{1}{2}\epsilon^{\alpha \beta} \delta_{ij}
- K P_{ij} \epsilon^{\alpha \gamma} [{\cal H}_{ij}]_\gamma^\beta \, ,
\label{epsD}
\end{equation}
or in short ${\widehat{\cal D}} = \epsilon {\cal D}$~:
\begin{equation}
\widehat{\cal D} = \widehat{\cal D}_0 - K \widehat{\cal H} 
\end{equation}
where $\widehat{\cal D}_0 = 
\frac{1}{2} \epsilon \times \mathbbm{1}$ is a deterministic
part, and $\widehat{\cal H}$ 
is a random part which consists of two-by-two 
matrices $\widehat{\cal H}_{ij} = 
\epsilon {\cal H}_{ij}$ (\ref{Hm}) located
at nonvanishing positions of the adjacency matrix
$P_{ij}$ which correspond to pairs of neighboring
triangles. $P_{ij}$ is an off-diagonal random matrix in 
the $ij$ indices.

One can show that $\widehat{\cal D}$ is antisymmetric
under the change of pairs of indices
\begin{equation}
\widehat{\cal D}_{ij}^{\alpha\beta} =
- \widehat{\cal D}_{ji}^{\beta\alpha} \, ,
\end{equation}
and hence
${\rm Pf}^2 \, \widehat{\cal D}  = {\rm Det} \, \widehat{\cal D} =
{\rm Det} \, \cal D$.
For each triangulation individually, the integral over fermions in
(\ref{Z}) yields Pfaffian of the matrix $\widehat{\cal D}$.
Thus for Majorana fermions on a two-dimensional triangulation
the partition function (\ref{Z}) is a sum of Pfaffians of the
Dirac-Wilson operator
\begin{equation}
{\cal Z} = \sum_{T \in {\cal T}} {\cal Z}_T =
\sum_{T \in {\cal T}} {\rm Pf} \, \widehat{\cal D}_T =
\sum_{T \in {\cal T}} {\rm Det}^{1/2} {\cal D}_T
\, .
\label{D}
\end{equation}
In the last step we have used the 
inequality ${\rm Pf} \, {\cal D}_T>0$,
which can be proven by the hopping parameter expansion.
The consistency condition (\ref{sp}) turns out
to be essential in the proof.
Namely, one shows that the Pffafian is
represented as a sum over loop configurations
each of which contributes a positive factor
if the condition (\ref{sp}) is met \cite{bjk,bbjkpp}.

Using this expansion one can also
establish the equivalence between the partition function
${\cal Z}_T$ and the partition function
of the nearest neighbor Ising model
with spins $\sigma_{i_*}$ located at the
vertices $i_*$ of $T$
\begin{equation}
Z_T = \sum_{\{\sigma_*\}_T} e^{-\beta E_T}
\label{iz}
\end{equation}
where
\begin{equation}
E_T = - \!\!\!\! \sum_{(i_*j_*)\in T} (\sigma_{i_*} \sigma_{j_*}-1) \, .
\end{equation}
The partition functions $Z_T$ (\ref{iz}) and ${\cal Z}_T$ (\ref{Z})
are equal for
\begin{equation}
K = \frac{e^{-2\beta}}{\sqrt{3}} \, .
\label{kb}
\end{equation}
In the derivation of this equivalence
one identifies loop configurations,
arising in the hopping expansion of ${\cal Z}_T$
with  domain walls of the Ising model. Again, the
consistency condition (\ref{sp}) plays the crucial role here.
For a non-spherical triangulation
one has to carefully treat topological effects related
to the existence of non-contractable loops which
may give a negative contribution for antiperiodic
boundary conditions. One can get rid of all negative
contributions performing the GSO projection
that is summing over all spin structures of the manifold \cite{gsw}.
This and another topological issues are discussed
elsewhere \cite{bbjkpp}. Here we will restrict ourselves to
spherical triangulations for which
we automatically have ${\cal Z}_T = Z_T$
for each triangulation $T \in {\cal T}$  and
hence also for the sum over all
triangulations in ${\cal T}$
\begin{equation}
{\cal Z} = Z \equiv \sum_{T \in {\cal T}} Z_T \, .
\end{equation}
The critical temperature of the Ising model for the partition
function $Z$ is known analytically to be
$\beta= \frac{1}{2} \ln \frac{131}{85} \approx 0.2162730$ \cite{bj}.
Translating the critical temperature
to the hopping parameter (\ref{kb})
we obtain the following critical value
\begin{equation}
K_{cr} = \frac{85\sqrt{3}}{393} \approx 0.3746166
\label{crit}
\end{equation}
for which fermions become massless. Another interesting point
which can be deduced from the equation (\ref{kb}) is that
the value of the hopping parameter
$K_0=1/\sqrt{3}$ corresponds to $\beta=0$ which is
the border between the ferromagnetic and
antiferromagnetic regimes. For $\beta<0$
one expects frustration in the Ising model on a triangulation
and hence that the lowest energy state is highly
degenerated. As we will see below, the behavior of
the spectrum of the Dirac-Wilson is also  
sensitive to passing over this border. 
First, we will however restrict $K$ to the 
'ferromagnetic' range $[0,K_0]$.

The equivalence of the partition functions
$Z_T$ and ${\cal Z}_T$ may be used to relate
the average energy $\overline{E}_T$
of the Ising model on a triangulation $T$ to eigenvalues
of the Dirac-Wilson operator. Differentiating both sides
of (\ref{D}) with respect to $\beta$
we obtain
\begin{equation}
\overline{e}_T = \frac{\overline{E}_T}{N} =
-\frac{1}{N} \, \frac{\partial}{\partial \beta} \ln Z_T =
1 - \frac{\sum_a \lambda_a^{-1}}{2N} \, ,
\label{el}
\end{equation}
where $\lambda_a$ are eigenvalues of the Dirac-Wilson operator
${\cal D}_T$.
For our choice of the representation of gamma matrices,
${\cal D}_T$ is a real matrix. Its spectrum consists of either
real eigenvalues or of pairs of complex conjugates. Thus
the sum on the right hand side of (\ref{el}) is a real number.
Similarly, the fluctuations of the Ising energy on the triangulation
$T$ are given by
\begin{equation}
\sigma^2_T = \frac{\overline{E^2}_T - \overline{E}_T^2}{N} =
\frac{1}{N} \, \frac{\partial^2}{\partial \beta^2} \ln Z_T =
- \frac{\sum_a \lambda_a^{-2}}{2N} + \frac{\sum_a \lambda_a^{-1}}{N} \, .
\label{sl}
\end{equation}
Averaging over triangulations we obtain the energy density
and heat-capacity of the Ising model coupled to gravity
calculated in terms of the eigenvalues of the Dirac-Wilson operator
\begin{equation}
e = \langle \overline{e}_T \rangle =
1 - \bigg\langle  \frac{1}{2N} \sum_a \lambda_a^{-1} \bigg\rangle \, ,
\end{equation}
\begin{equation}
c_v = \beta^2 \langle \sigma^2_T \rangle \, =
\beta^2\left\{ - \bigg\langle \frac{1}{2N}
\sum_a \lambda_a^{-2} \bigg\rangle  +
\bigg\langle \frac{1}{N} \sum_a \lambda_a^{-1} \bigg\rangle \right\}\, .
\end{equation}
The equivalence of the models can also be very useful
in MC simulations of the model. To show this, let us
compare three numerical experiments in which (a)
the Ising model is used to generate
triangulations and to measure the Ising energy and heat capacity;
(b) the Ising model is used
as a generator of triangulations but measurements
are carried out using the fermion field;
(c) the fermionic model is used both to generate
triangulations and to perform measurements.
\begin{figure}
\begin{center}
\psfrag{xx}{$\beta$}
\psfrag{yy}{$c_v$}
\includegraphics[width=8cm]{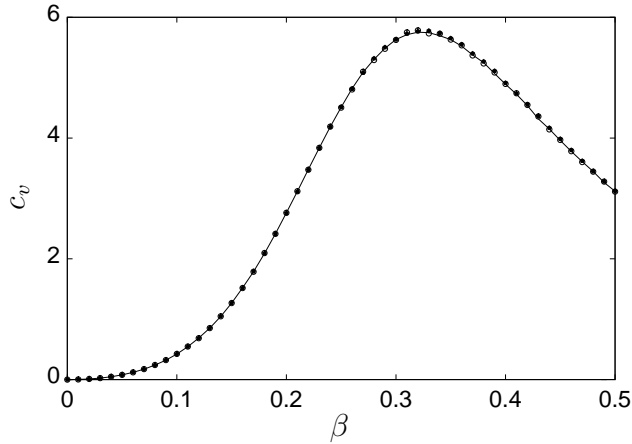}
\caption{\label{Fabc}Heat capacity $c_v(\beta)$
for the system with $N=16$ triangles for the three
cases discussed in the text~: (a) ising-ising (line);
(b) ising-fermion (filled symbols); (c) fermion-fermion
(empty symbols). The three methods give the same results within
the error bars, which are here of order of the
size of the symbols used.}
\end{center}
\end{figure}
As shown in fig.\ref{Fabc} the three methods yield
exactly the same results. The methods differ, however,
significantly in the CPU time needed to generate results
of the same quality. The first difference comes from the
configuration generator which is much faster for the Ising
model than for the fermionic determinant. In the latter case,
to calculate a Metropolis weight for a single local change of triangulation,
{\em i.e.} a flip of one link on the triangulation, requires the
recomputation of the determinant of the Dirac-Wilson operator on
the modified lattice. This is a tedious task for which the number
of operations grows with the third
power of the system size $N$. Thus one expects that the time of
a sweep through the lattice grows as $N^4$ for the fermionic
configurations generator. One sweep for the Ising model, which
consists of a sweep of local updates of Ising spins, a fixed number of
Wolff cluster updates,
and a sweep of local changes of triangulation,
lasts in CPU units roughly proportionally to the system size $N$.
Thus, the fermionic algorithm is competitive with the Ising generator
only for very small lattices.
As far as measurements are concerned the
situation is more complex. For example, one cannot determine
the spectrum of the Dirac-Wilson operator using only the Ising
spins. One can, however, do the opposite.
For a given lattice, the time of calculating all eigenvalues
of the Dirac-Wilson operator is proportional to $N^3$. Having
done this, one is able for this triangulation
to exactly calculate the Ising energy (\ref{el}) and
it higher moments (\ref{sl}) without statistical fluctuations.
If one instead used the Ising model, one has to
sample Ising configuration many times to reduce the error.
In general, the cost of a single measurement of the energy is
proportional to $N$. The error of the single measurement
of the energy density decreases like $1/\sqrt{N}$.
Summarizing, we expect the CPU time to measure energy
with a given precision to grow as $\sqrt{N}$. The CPU time
grows rapidly with the order for measurements
of higher moments of energy.

In order to obtain the data the quality presented in
fig. \ref{Fabc} for $N=16$,
the methods discussed above required
(a) 1000 CPU min. (b) 6 CPU min. and (c)
100 min. on the computer Alpha XP1000/EV6/500 MHz.

\section*{Spectrum of the Dirac-Wilson operator}

\begin{figure}
\begin{center}
\psfrag{xx}{Re}
\psfrag{yy}{Im}
\includegraphics[width=8cm]{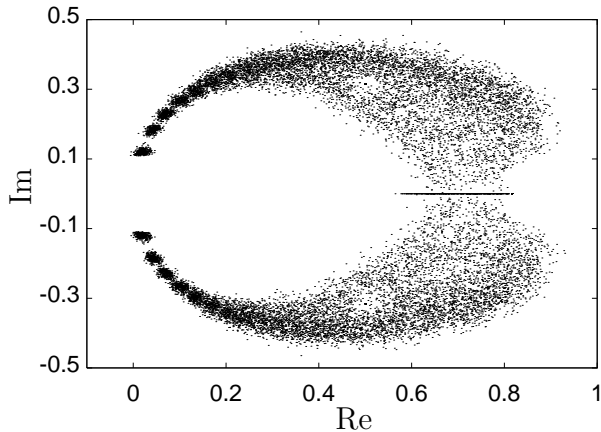}
\caption{\label{Eig} The distribution of eigenvalues
$\lambda$ of the Dirac-Wilson operator for $N=64$ and
$K=0.364$ on random lattice.}
\end{center}
\end{figure}

In the production runs we use the method (b), which relies
on generating triangulations from the partition
function of the Ising model. At each measurement
we ignore the values of the Ising spins
and we assign frames $e_i$ and
$s_{ij}$-signs to the triangulation to reconstruct the
Dirac-Wilson operator (\ref{Hm},\ref{dwo}).

A typical spectrum of the Dirac-Wilson operator on random
triangulation is shown in fig.\ref{Eig}. The main effect
on the spectrum of changing the hopping parameter $K$ is
to rescale it around the point $(\frac{1}{2},0)$.
The positions of the two claw-shaped ends of the
spectrum move with $K$. One can find a value of $K$ for which
the ends lie closest  the origin $(0,0)$. This value can be 
treated as a pseudo-critical value $K_*$ for which
the mass of the fermion excitation is minimal.
For $K< K_*$ the origin $(0,0)$ lies outside the claws,
while for $K > K_*$ inside. In fact, this is the main difference
between the two regimes, since
beside the scaling factor the shape of the spectrum
is almost constant.

The claw-shaped ends of the pseudo-critical spectra
successively approach each other when the system size $N$ 
is increased. They eventually close 
entirely at the origin $(0,0)$ for infinite $N$, signaling
the occurrence of massless excitations.

One can alternatively study the spectrum of the operator
$\widehat{\cal D} = \epsilon {\cal D}$. In fact, this operator
is closer related, in spirit, to Majorana fermions (\ref{epsD})
due to the presence of the charge conjugation matrix $\epsilon$.
Since the matrix $\widehat{\cal D}$ is antisymmetric, and it is
real in our representation, its
spectrum is purely imaginary. Thus,
the eigenvalue density of this operator is one dimensional~:
\begin{equation}
\widehat{\rho}(x) = \lim_{N\rightarrow \infty} \frac{1}{N} \left\langle
\sum_{\widehat{\lambda}} \delta(x-i\widehat{\lambda} \, ) \right\rangle \, .
\end{equation}
The spectrum is symmetric $\widehat{\rho}(x) = \widehat{\rho}(-x)$. 
We will constrain ourselves to the positive branch.
For each triangulation eigenvalues of the positive part of
the spectrum can be ordered~: 
$\lambda_0 \le \lambda_1 \le \lambda_2 \le \dots $.
Collecting separately histograms for the lowest, 
second lowest, third lowest eigenvalues {\em e.t.c.}
one obtains distributions $\widehat{\rho}_j(x)$ 
of the $j$-th eigenvalue. Of course, by construction~: 
$\widehat{\rho}(x) = \sum_j \widehat{\rho}_j(x)$.

\begin{figure}
\begin{center}
\psfrag{xx}{$\widehat{\lambda}$}
\psfrag{yy}{$\widehat{\rho}$}
\psfrag{xy}{ }
\psfrag{etyk}{a}
\includegraphics[width=6cm]{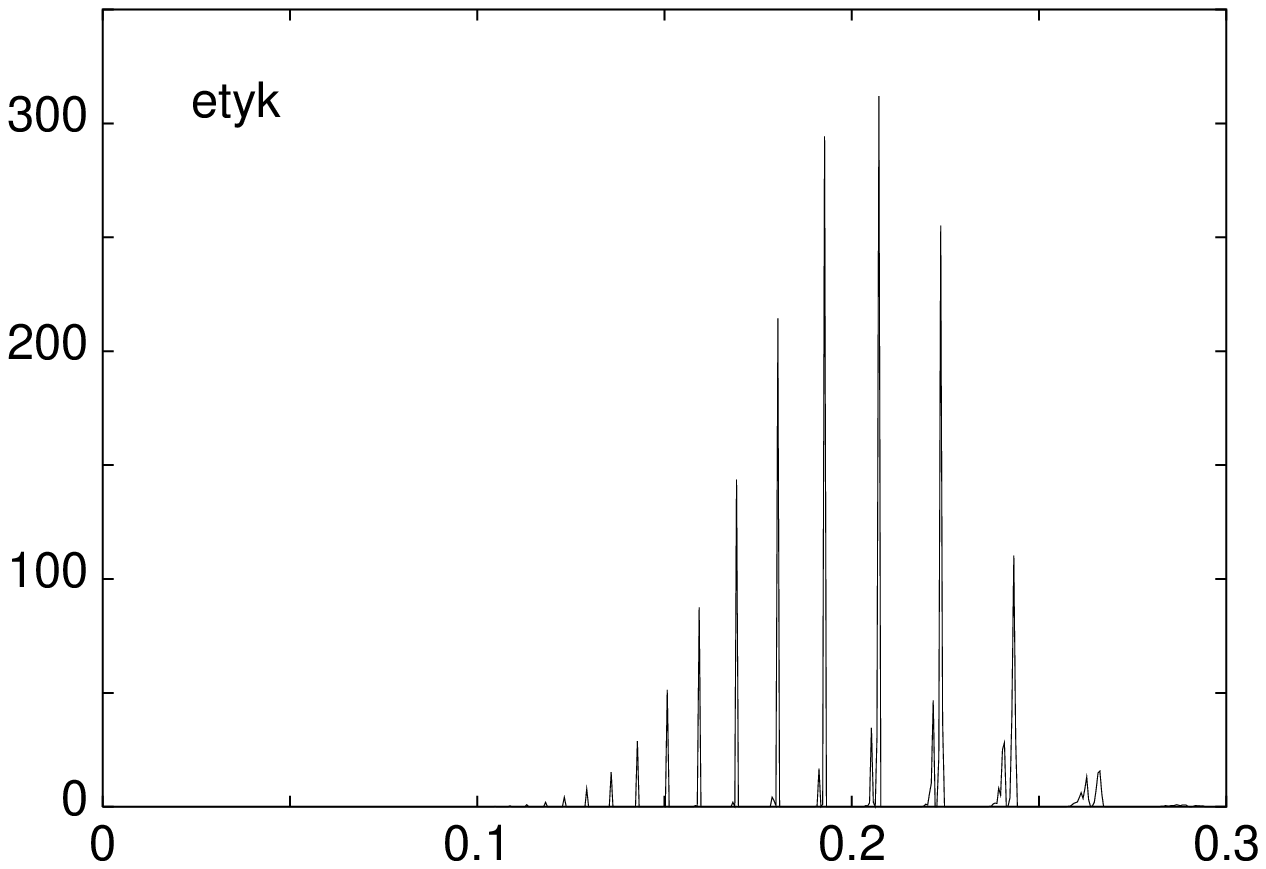}
\psfrag{etyk}{b}
\includegraphics[width=6cm]{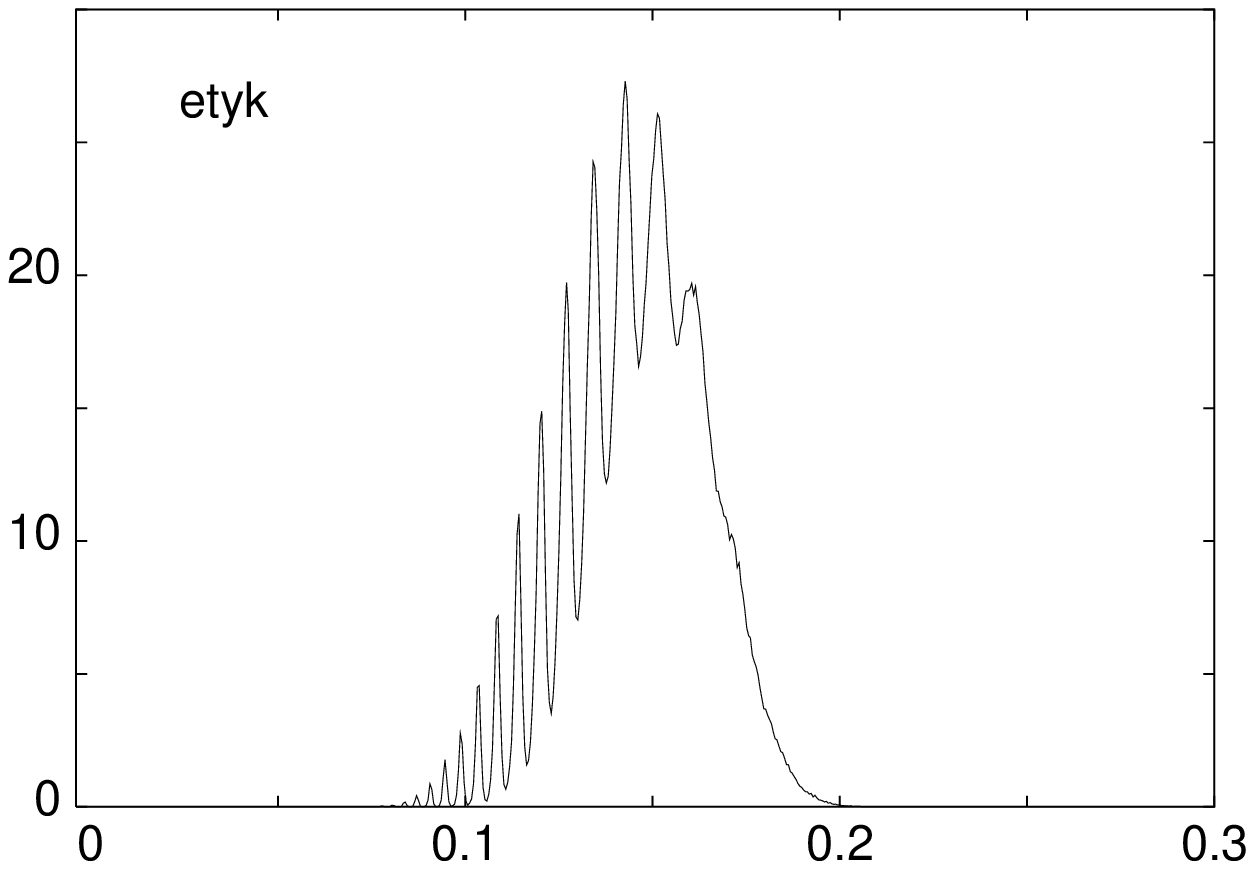} \\
\psfrag{etyk}{c}
\includegraphics[width=6cm]{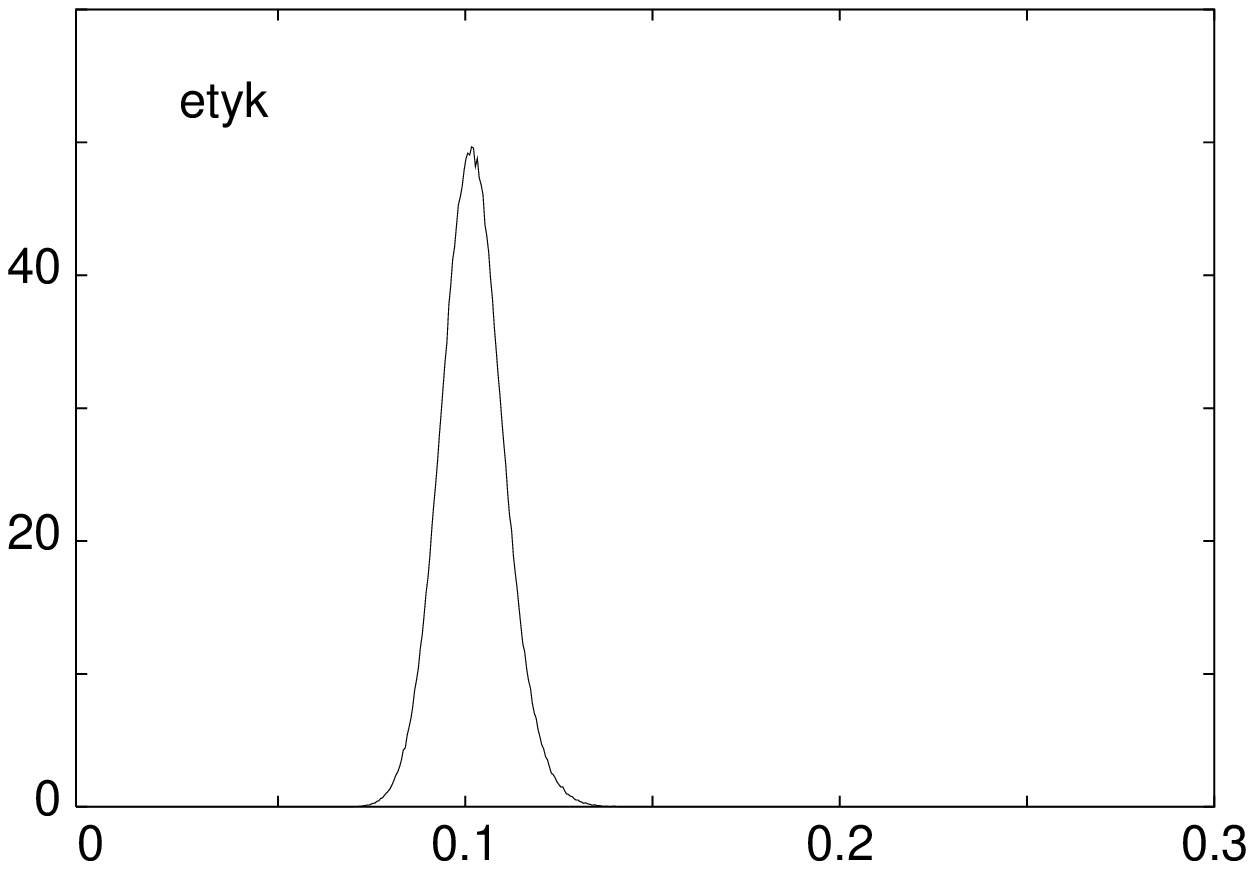} 
\psfrag{etyk}{d}
\includegraphics[width=6cm]{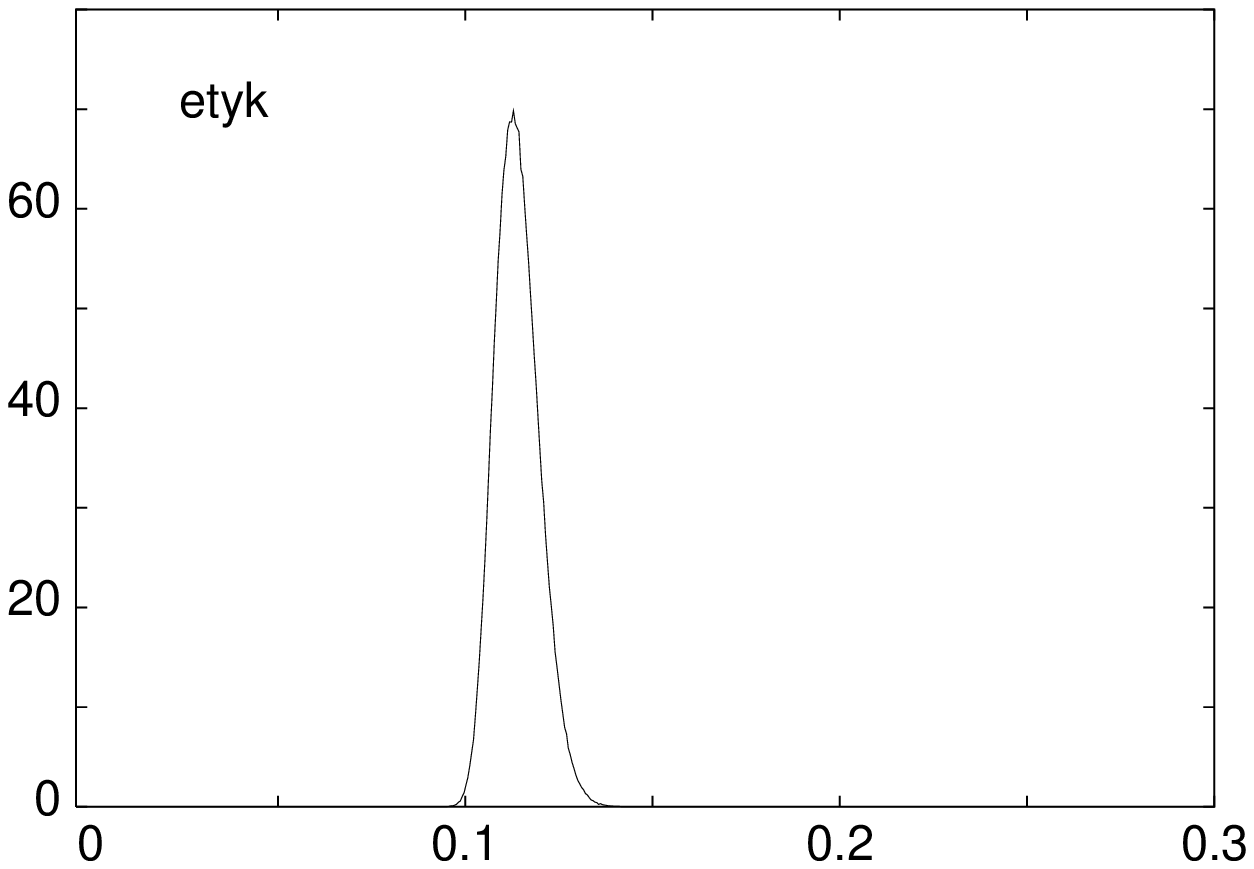}
\caption{\label{rho0} Evolution of the shape of the probability
distribution $\widehat{\rho}_0(x)$ 
of the lowest eigenvalue of the operator
$\epsilon {\cal D}$ for $N=64$
as a function of the hopping parameter $K$~:
(a) $K=1/\sqrt{3} = 0.5774$,
(b) $K=1/\sqrt{3} e^{-0.2} = 0.4727$,
(c) $K=1/\sqrt{3} e^{-0.4} =0.3870$,
(d) $K=1/\sqrt{3} e^{-0.6} =0.3169$.
Each histogram presented in the figure contains $7 \times 10^5$ counts. 
The bin size is $5 \times 10^{-4}$. The histograms are normalized
to have area one.}
\end{center}
\end{figure}

We studied numerically the dependence of $\widehat{\rho}_j(x)$ 
on the hopping parameter. In figure \ref{rho0}
is shown the distribution 
$\widehat{\rho}_0(x)$ for different values of $K$.
One can see that it is
discrete for $K_0=1/\sqrt{3}$ consisting of
separate narrow peaks. This may appear surprising at first glance.
However, when $K$ becomes smaller than $K_0$ this spectrum 
becomes continuous~: it gradually changes when $\Delta = K_0 - K$ grows
to eventually become a smooth bell-shaped distribution (fig.\ref{rho0}).
As we shall see below the discreteness of the
spectrum at $K_0=1/\sqrt{3}$ is not a finite size effect.
We made the following experiment. We generated 
a quenched ensemble of random triangulations by ignoring
the coupling of fermions to gravity. It is an ensemble
of triangulations for pure gravity. Then, 
for each triangulation from this ensemble, we calculated the
operator (\ref{epsD}) and determined its lowest eigenvalues  
for different values of $K$. One can expect that
outside the critical region, where fermions are massive,
the approximation should not significantly affect 
the spectrum of the model. Indeed we checked numerically
for a few values of $K$ that it is practically impossible
to distinguish between the spectrum for the quenched and the full model.

\begin{figure}
\begin{center}
\psfrag{xx}{$K$}
\psfrag{yy}{$\lambda$}
\psfrag{etyk}{a}
\includegraphics[width=6cm]{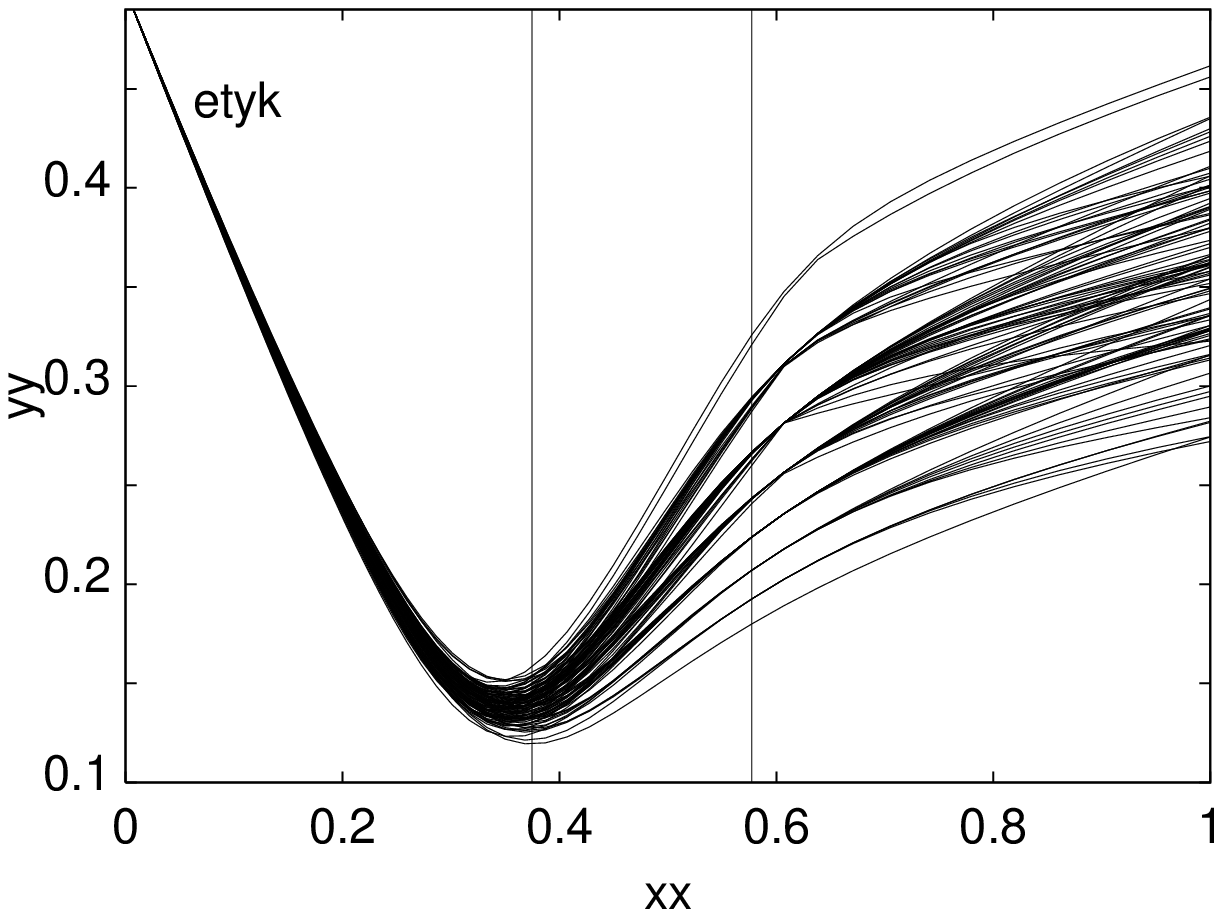}
\psfrag{etyk}{b}
\includegraphics[width=6cm]{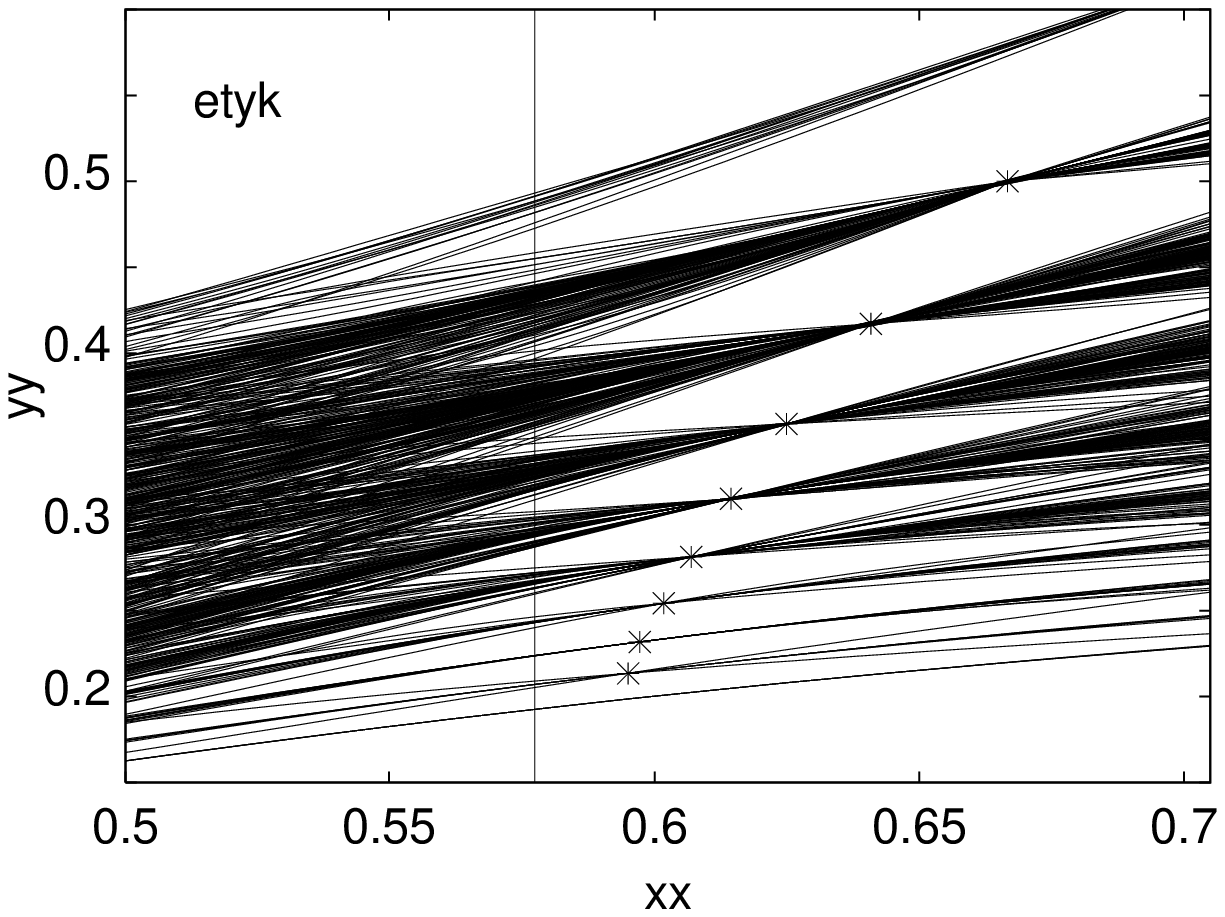} 
\caption{\label{bundle} (a) Evolution of the 
lowest eigenvalue of the operator (\ref{epsD}) with the hopping 
term, calculated on an ensemble of given triangulations with $N=32$
triangles. (b) Evolution of seven lowest eigenvalues of the 
operator (\ref{epsD}). The cross-points of the bundles in the 
'antiferromagnetic' phase are seen. The cross-points can be 
numbered  by successive integers from $6$ to $13$
corresponding to the length $q$ of the elementary 
loops, as discussed in the text.}

\end{center}
\end{figure}

The results of the quenched experiment are presented in figure
\ref{bundle} where one can see lowest eigenvalues 
of the operator (\ref{epsD}) for different $K$.
The 'ferromagnetic' region corresponds to the interval
$0\le K < 1/\sqrt{3} = 0.5774$, and the 'antiferromagnetic'
one $K > 1/\sqrt{3}$. Universal properties of 
the model are encoded in the behavior of the spectrum
around the critical value $K_{cr}=0.3746$ (\ref{crit}), 
lying deep inside the 'ferromagnetic' phase,
where the eigenvalue bundle has a dip. 
For $K \rightarrow 0$ only the 
deterministic part 
$\frac{1}{2}\epsilon^{\alpha \beta} \delta_{ij}$
of the operator (\ref{epsD}) survives.  

The bundles of eigenvalues have an interesting property,
The lines of the bundle cross at some points
in the 'antiferromagnetic' phase (see fig. \ref{bundle}).
The meaning of a cross-point is that, for the corresponding value
of $K$, the operator (\ref{epsD}) has a common identical eigenvalue 
for many different triangulations. The reason why it happens
is that this eigenvalue is entirely related to the existence of
elementary loops of the length $q$ on the triangulation 
and not to the whole random structure of the triangulation,
as it generically takes place.

In order to understand the mechanism of the
occurrence of the cross-points, 
consider an elementary loop on the dual lattice of
length $q$ consisting of vertices $i_1,i_2,\dots,i_q$.
One can show that the corresponding submatrix 
$2q\times 2q$ of (\ref{epsD}) built of the two-by-two 
blocks at the $(i_1,i_2,\dots,i_q) \times (i_1,i_2,\dots,i_q)$ 
positions has for some $K_q$ an eigenvalue $\lambda_q$ which 
depends only on $q$. 
The pair $(K_q,\lambda_q)$ corresponds to
a cross-point in the figure (\ref{bundle}).
Furthermore, it turns out that for $K=K_q$ and $\lambda=\lambda_q$ 
the entire $2q$ rows for $i_1,i_2,\dots i_q$ of the matrix
$\widehat{\cal D} - \lambda_q \mathbbm{1}$ are linearly dependent.
This means that $\lambda_q$ is not an eigenvalue 
of the submatrix but also of the whole matrix 
$\widehat{\cal D}$ for $K=K_q$ (\ref{epsD}). The pairs $(K_q,\lambda_q)$
correspond to the cross-points of eigenvalues bundles 
in fig. \ref{bundle}. They
can be numbered by $q$. We found 
$(K_q,\lambda_q)$ to be  $(2/\sqrt{3},3/2)$,
$(\sqrt{2}/\sqrt{3},\sqrt{3}/2)$,
$((\sqrt{5}-1)/\sqrt{3},\sqrt{15-6\sqrt{5}}/2)$, $(2/3,1/2)$
for $q=3,4,5,6$, respectively. Positions of the cross-points
do not change with the lattice size.
This structure of the cross-points will be discussed 
in more detail elsewhere. Here we only want to mention some 
features of this structure. All the points lie in 
'antiferromagnetic' phase. It is seen in 
fig.\ref{bundle} that for $q \rightarrow \infty$
the points approach the border to the 'ferromagnetic' phase 
$K_q \rightarrow K_0$ and that the corresponding 
eigenvalues $\lambda_q$ decrease. Moreover the probability
of encountering an elementary loop of length $q$ on random
triangulation falls off exponentially \cite{kkm} which means
that there are very few loops with high $q$. 
Taking all these facts into account one may expect that 
the presence of discrete line of cross-points which approaches
$K_0$ shall dominate the shape of the spectra for 
the lowest eigenvalues at $K_0$,
leading in particular to the appearance of 
discrete peaks also in the limit $N\rightarrow \infty$.
\begin{figure}
\begin{center}
\psfrag{etyk}{a}
\includegraphics[width=6cm]{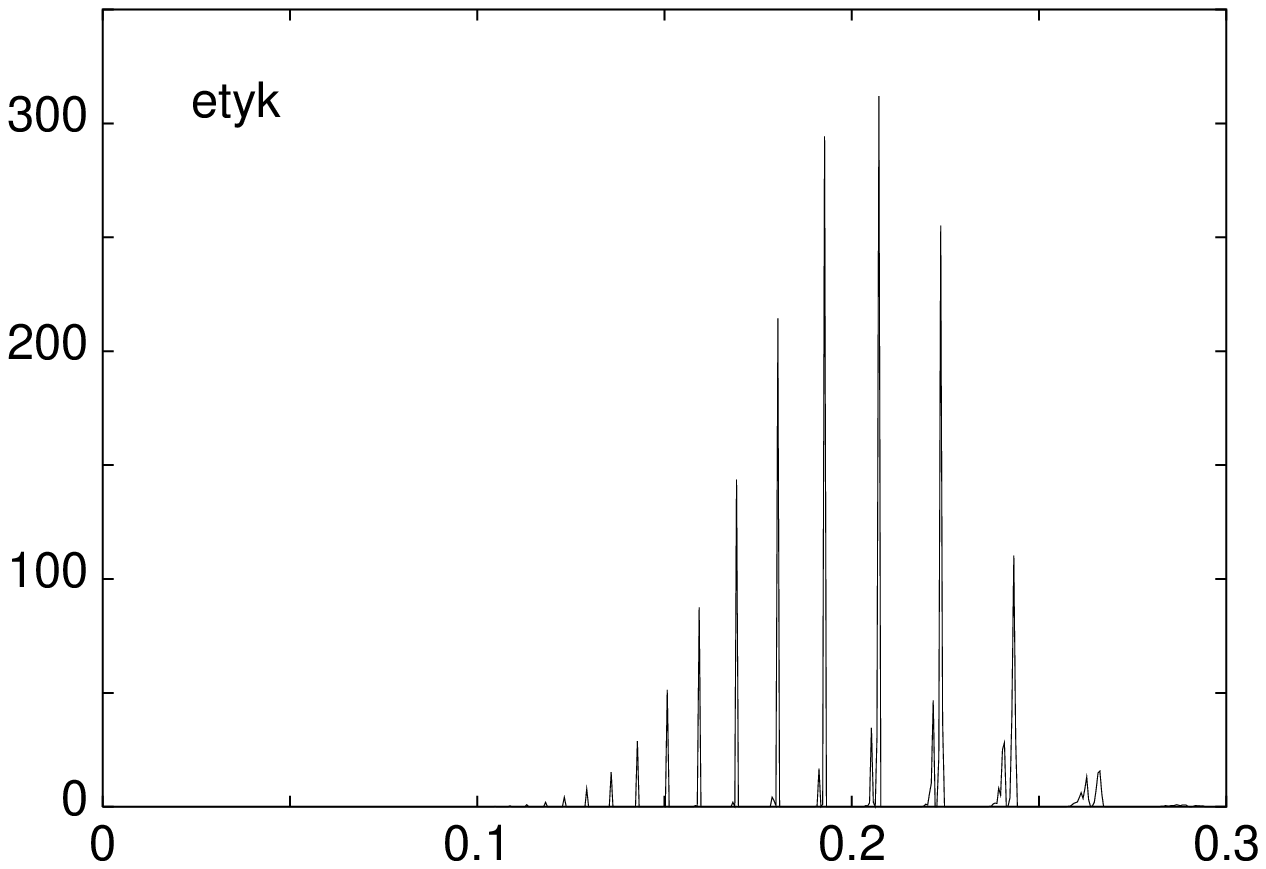}
\psfrag{etyk}{b}
\includegraphics[width=6cm]{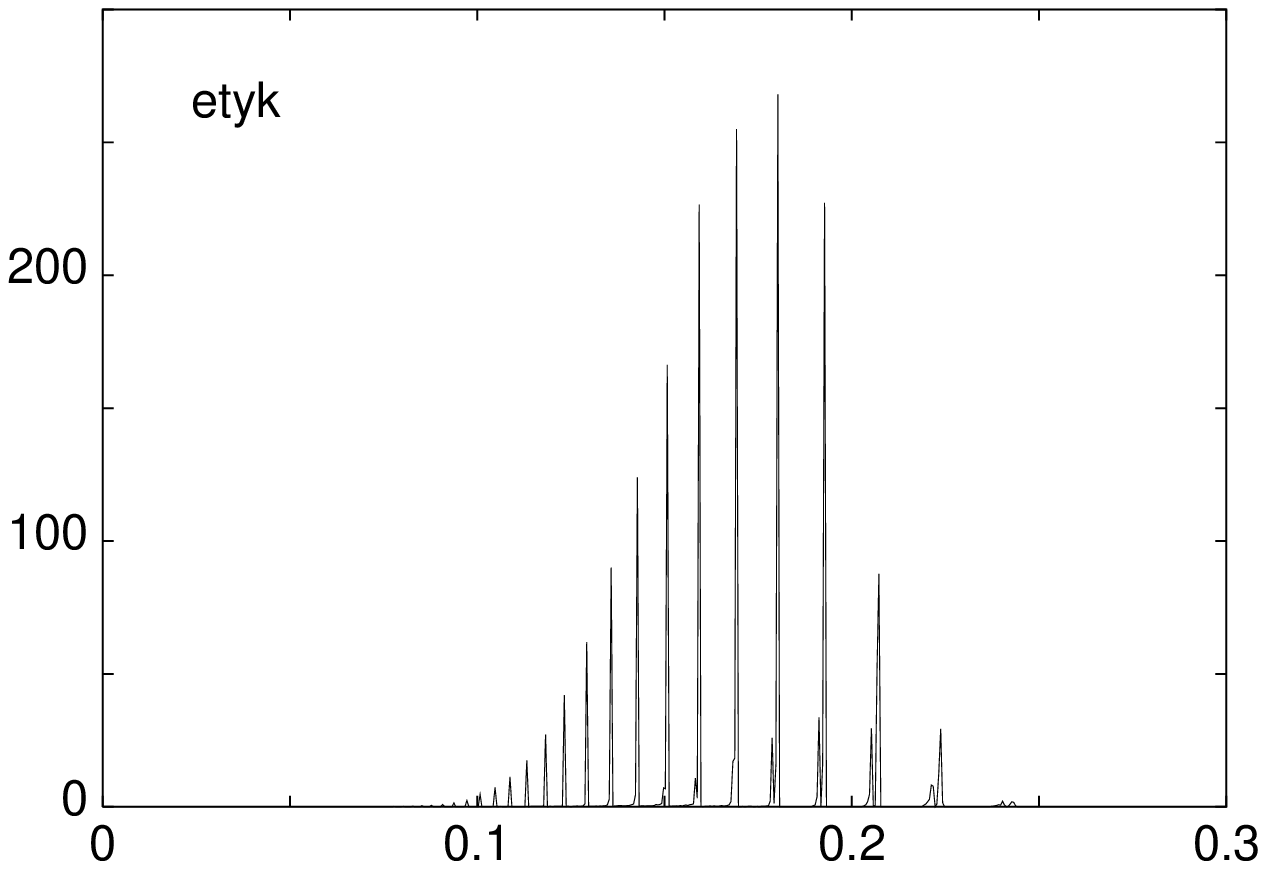} \\
\psfrag{etyk}{c}
\includegraphics[width=6cm]{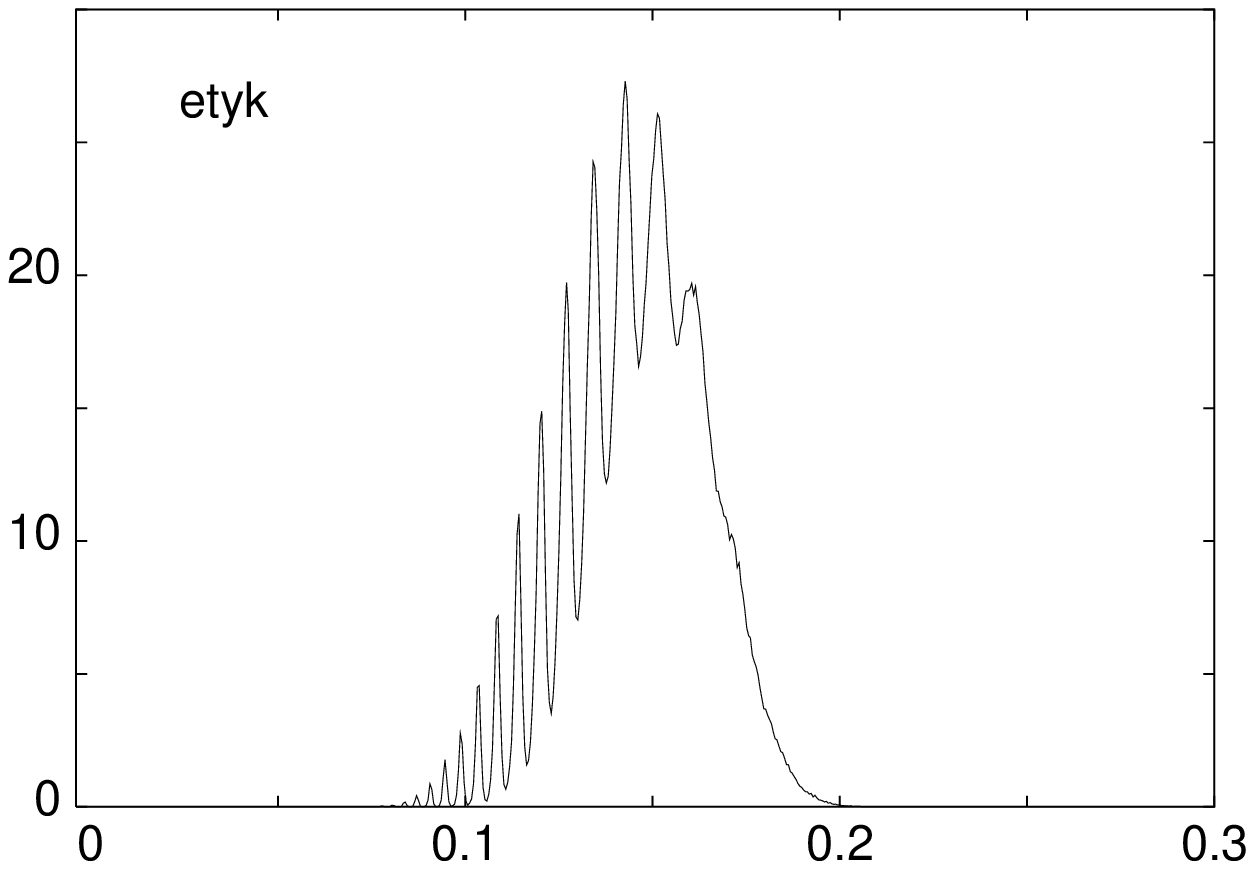} 
\psfrag{etyk}{d}
\includegraphics[width=6cm]{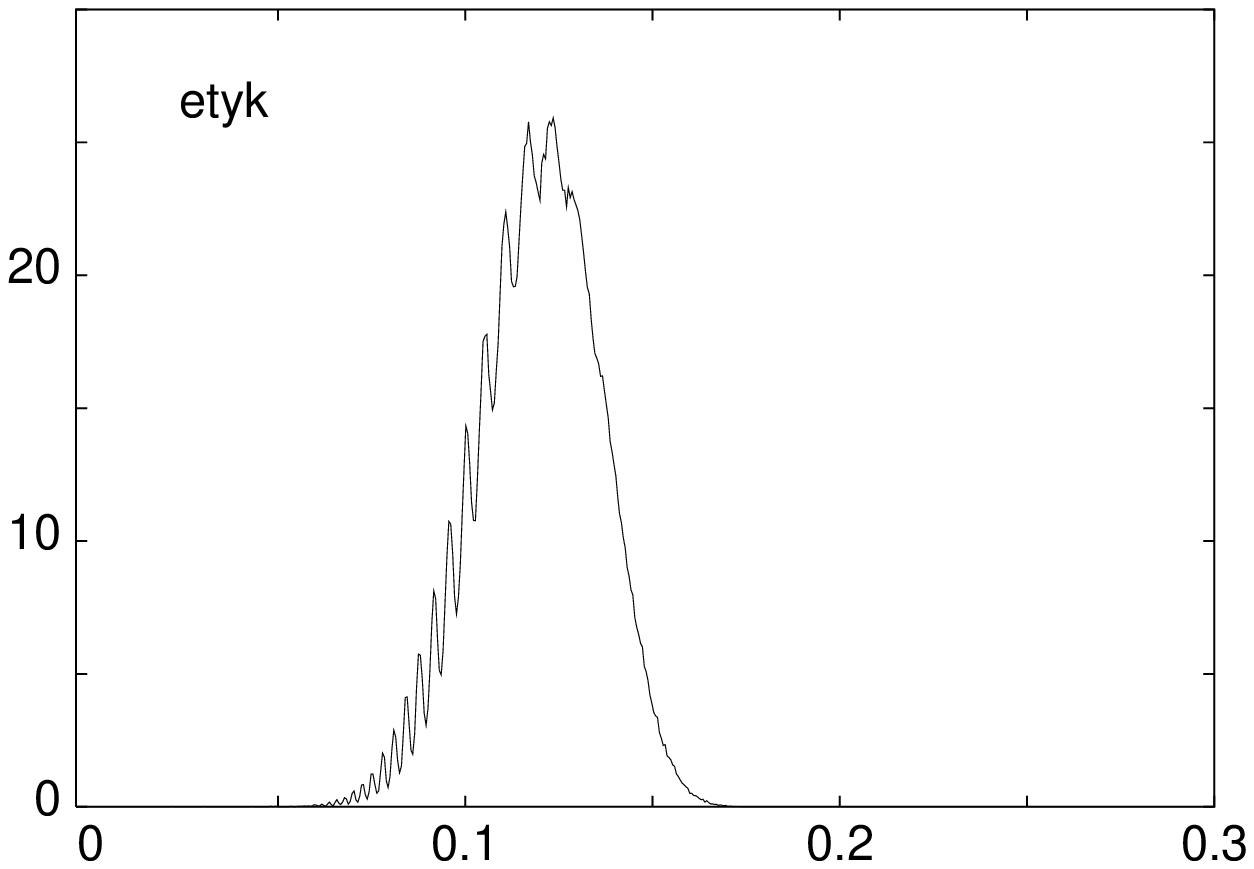}
\caption{\label{osc} The probability distribution
$\widehat{\rho}_0(x)$ for (a) $K_0=1/\sqrt{3}$ ,
$N=64$, (b) $K_0 = 1/\sqrt{3}$ , $N=128$.
(c) $K=1/\sqrt{3} e^{-0.2}=0.4727$, $N=64$ 
(d) $K=1/\sqrt{3} e^{-0.2}=0.4727$, $N=128$.
The amplitude of the oscillations in the bottom row decreases with $N$. 
The histograms (a) and (b) contain $2\times 10^5$ counts each,
(c) $6\times 10^5$, (d) $7 \times 10^5$.
The bin size is $5 \times 10^{-4}$. }
\end{center}
\end{figure}

On the other hand, one also expects that
the discreteness of the spectrum disappears when $K$ goes deeper
into the ferromagnetic phase.
The crossover between the two regimes depends 
on some combination of $\Delta K = K_0 - K$ and the system size $N$.
For $\Delta K$ larger than zero, 
indeed the amplitude of the fluctuations 
decreases with the size as can be seen  in fig.\ref{osc}.
The fluctuations disappear in the limit $N\rightarrow \infty$ leaving
out a smooth distribution. 

\begin{figure}
\begin{center}
\psfrag{etyk}{a}
\includegraphics[width=6cm]{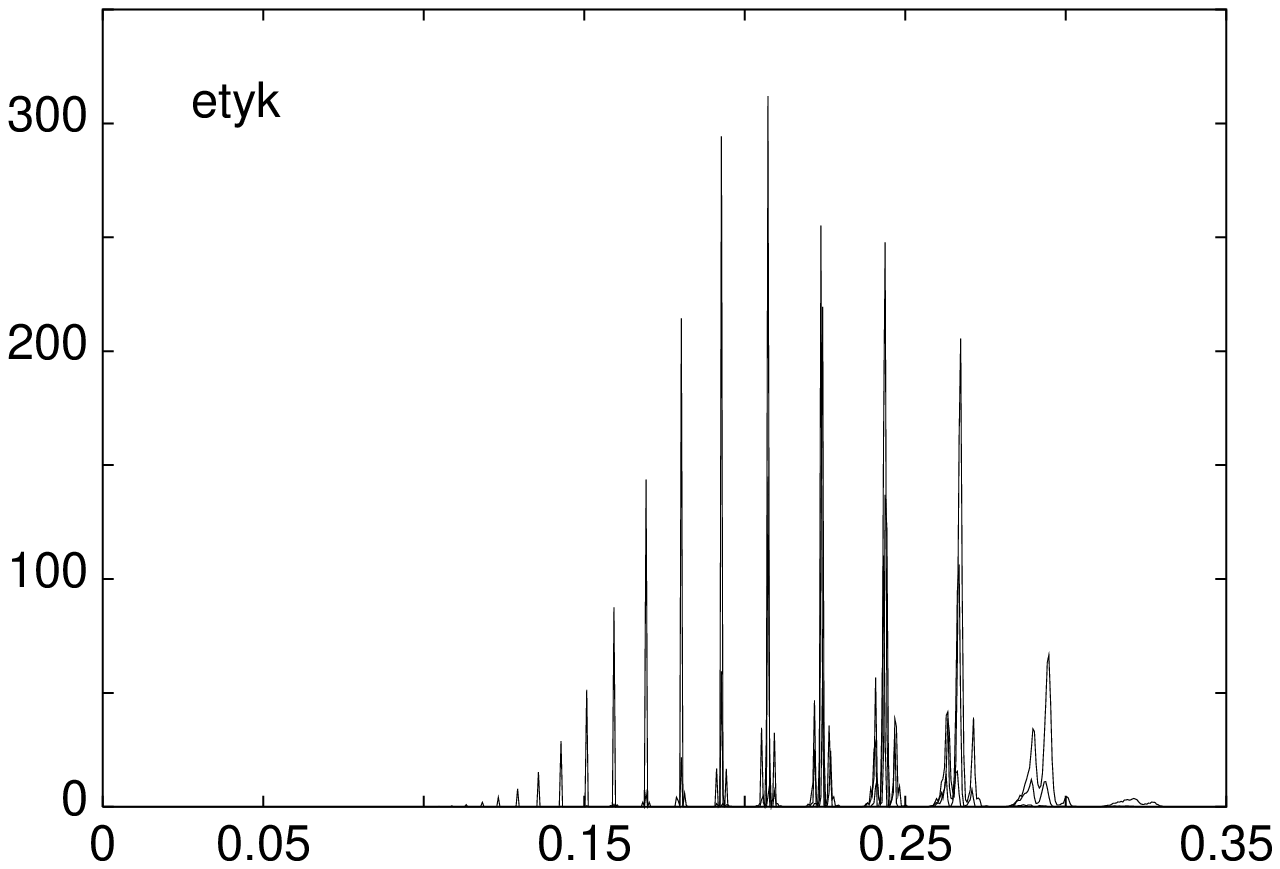}
\psfrag{etyk}{b}
\includegraphics[width=6cm]{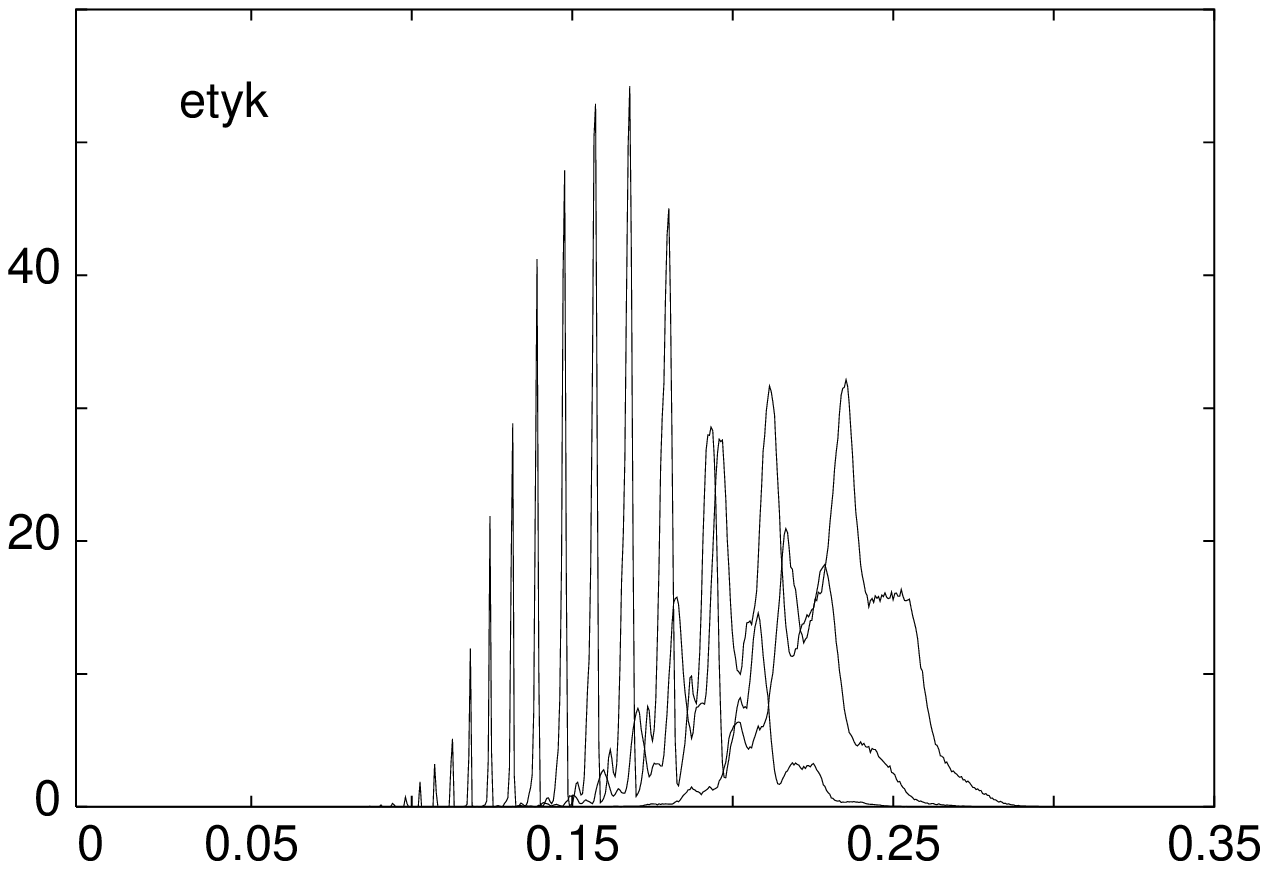} \\
\psfrag{etyk}{c}
\includegraphics[width=6cm]{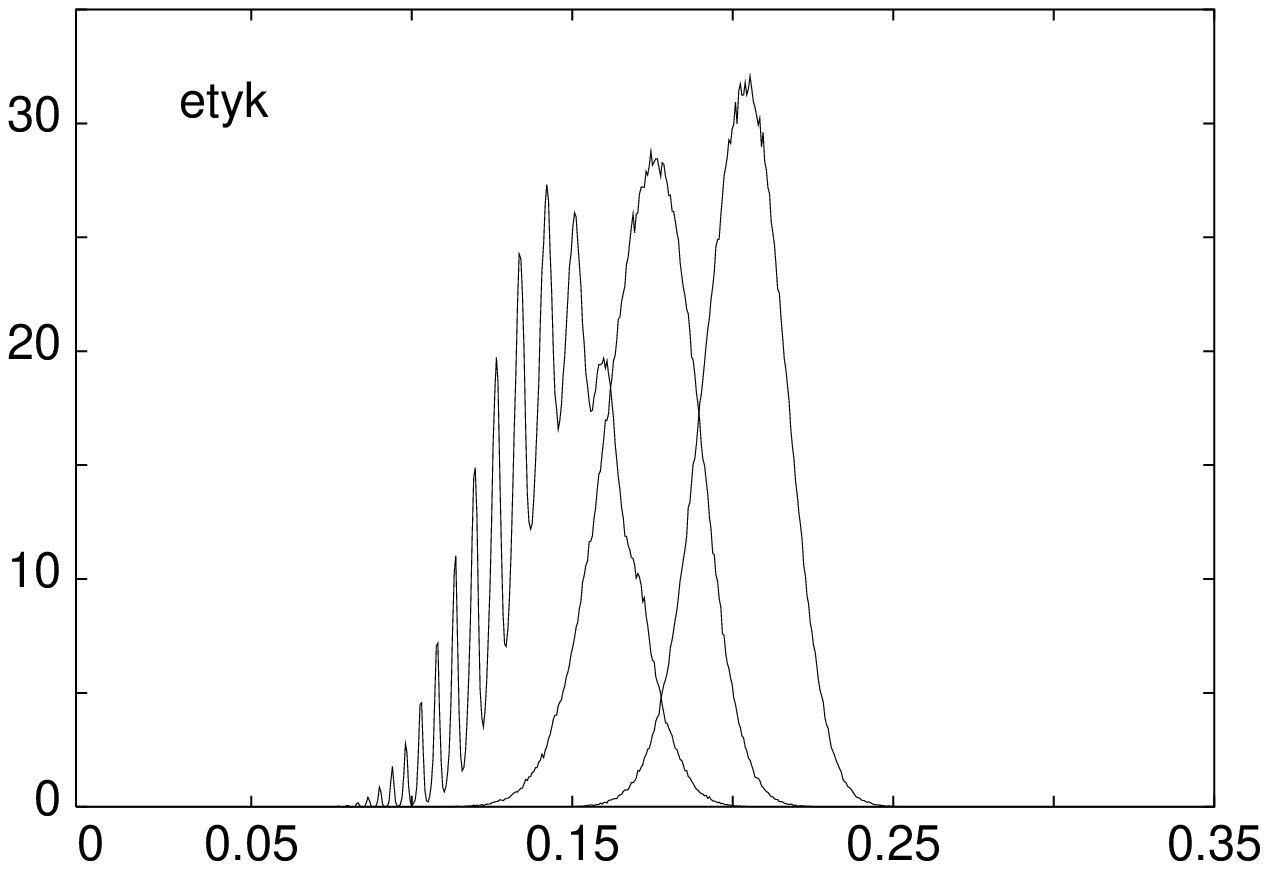} 
\psfrag{etyk}{d}
\includegraphics[width=6cm]{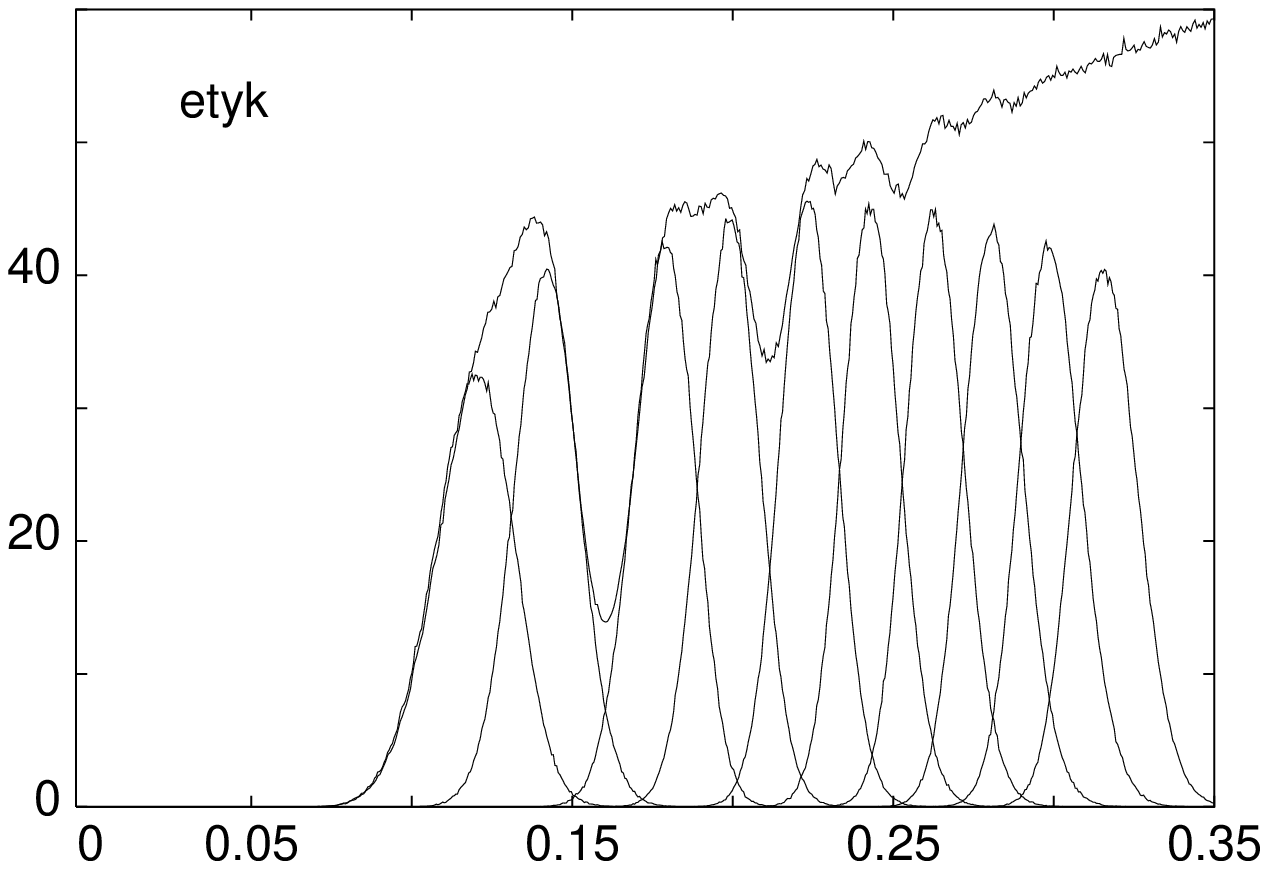}
\caption{\label{lev} The figures show probability distributions
$\widehat{\rho}_j(x)$ of the lowest eigenvalues for $N=64$ and 
(a) $K=1/\sqrt{3} = 0.5774$,
(b) $K=1/\sqrt{3} e^{-0.1} =0.5224$,
(c) $K=1/\sqrt{3} e^{-0.2} =0.4727$,
(d) $K=1/\sqrt{3} e^{-0.3} =0.4277$.
More precisely, each figure (a-c) contain
distributions for $j=0,1,2$, while the figure (d) for
$j=0,1,\dots,9$. 
Additionally, in the figure (d) the p.d.f. $\widehat{\rho}(x)$
is shown. Each histogram $\widehat{\rho}_j(x)$
contains $6 \times 10^5$ counts and has
the bin size is $5 \times 10^{-4}$, while
$\rho(x)$ contains $10^6$ counts, and has the bin
size $1.25 \times 10^{-3}$. 
}
\end{center}
\end{figure}

The spectra $\widehat{\rho}_j(x)$ 
of other eigenvalues $\lambda_j$, $j>0$, also exhibit
the same fluctuating pattern at $K_0$ as for $j=0$, however the
amplitude of the fluctuations decreases faster with
$\Delta K$ and $N$. As one can, for example, see in fig.\ref{lev}.c
the oscillations are absent 
in the spectrum of the third lowest eigenvalue $\rho_2(x)$ 
already for $K=0.4727$.

For $K$ far from $1/\sqrt{3}$ density distributions
are smooth. In this case, the situation resembles the one known 
for instance from the considerations of chiral matrix models~:
separate densities $\widehat{\rho}_j(x)$
are described by bell-shaped functions \cite{dn}. They sum
up $\widehat{\rho}(x) = \sum_j \widehat{\rho}_j(x)$
to a function with oscillations \cite{vz} like for example in
fig.\ref{lev}.d. It would be interesting to find 
a random matrix model which reproduces $\widehat{\rho}_j(x)$ and
$\widehat{\rho}(x)$ analytically.

Another interesting feature of the histograms
is the presence of singular peaks for which the number
of entries grow with the size of the lattice.
The peaks lie outside the range displayed in the figures.
We show in the appendix that such peaks 
are also present in the spectrum on the regular lattice.
In this case we calculated analytically that the height of the
peaks is logarithmically divergent in the lattice size.
Thus this seems to be a generic situation.

Let us come back to the universal
properties of the system of fermions interacting
with gravity. We will study now in more detail 
behavior of the spectrum 
at the critical point of the Ising model. More precisely
we shall be interested in the mass gap of the of the spectrum 
at $K$ close to $K_{cr}$.
We define the gap as the position of the center of mass 
for the distribution of the lowest eigenvalue 
$\widehat{\rho}_0(x)$. We denote it by 
$M = \int x \widehat{\rho}_0(x)$. We want to determine 
the dependence $M=M(K)$ on the hopping parameter $K$ 
for the given system size $N$. We do this 
numerically using the Lanczos 
algorithm\footnote{The Lanczos algorithm
\cite{lancz} is an iterative procedure
to calculate eigenvalues. It is frequently used
to approximately determine the lowest part of the
eigenvalue spectra of large matrices, for which
exact standard diagonalization algorithms would
require a too long time. In a single
iteration step the Lanczos algorithm finds one
approximated eigenvalue and improves quality of
the previously calculated ones.
As a rule it first produces the smallest and the largest
eigenvalues and then successively fills up the remaining
part of the spectrum,
The accuracy increases with the number of iterations.
We checked in our case using matrices of sizes up to $128$,
that when we keep the number $n$ of iteration
proportional to the size of the matrix $n=cN$ with,
$c=0.25$, the distribution of the lowest eigenvalue agrees
with the one obtained by an exact algorithm.}.

\begin{figure}
\begin{center}
\psfrag{xx}{$K$}
\psfrag{yy}{$M$}
\includegraphics[width=8cm]{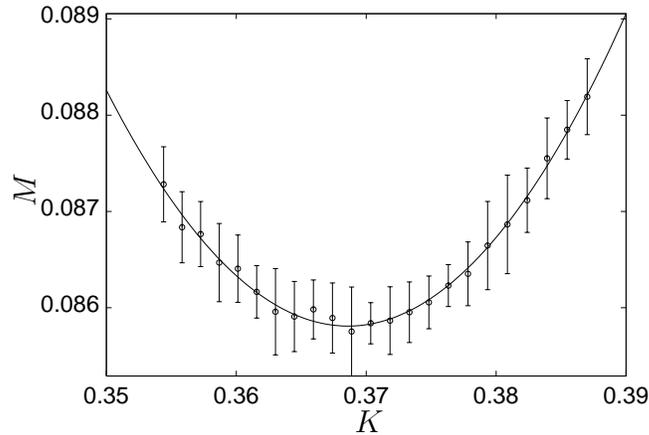}
\caption{\label{minK} The position of the center of mass is shown
for the distribution of the smallest eigenvalue of the
operator $\widehat{\cal D}$ for $N=96$
as a function of the hopping parameter $K$.}
\end{center}
\end{figure}

For given $N$ the function $M(K)$ 
has a minimum (see fig.\ref{minK}). The value
of the minimum $M_*$ plays the role of a mass gap,
while its position $K_*$ of a pseudo-critical
hopping parameter.
We determined $K_*$ and $M_*$ for different system sizes.
The results are collected in the table \ref{t1}.
\begin{table}
\begin{center}
\begin{tabular}{|c|c|c|}
\hline
$N$ & $K_*$ & $M_*$ \\
\hline
32 & 0.352(3) & 0.1395(5)  \\
48 & 0.360(2) &  0.1162(4) \\
64 & 0.364(2) &  0.1016(2) \\
96 & 0.368(2) &  0.0858(2) \\
128 & 0.370(1)& 0.0766(2) \\
192 & 0.372(2)& 0.0656(3) \\
256 & 0.372(2)& 0.0589(3) \\
384 & 0.374(2)& 0.0509(3) \\
512 & 0.374(2)& 0.0459(4) \\
768 & 0.375(3)&  0.0397(2) \\
1024 & 0.375(1)& 0.0359(3) \\
\hline
\end{tabular}
\end{center}
\caption{ \label{t1} Positions and values of the minima
of the function $M(K)$ representing the center of
mass of the distribution of the smallest eigenvalue
of the operator $\widehat{\cal D}$ for different system
sizes $N$.}
\end{table}
We fitted the data points to the following finite size
scaling formulas
\begin{equation}
M_* = \frac{b}{N^{\frac 1 {d_H}}}\left(1+\frac{t}{N}\right) \quad , \quad
K_{*} = K_{\infty} - \frac{a}{N^\kappa} \, .
\label{scaling}
\end{equation}

\begin{figure}
\begin{center}
\psfrag{xx}{$N$}
\psfrag{yy}{$M_*$}
\includegraphics[width=8cm]{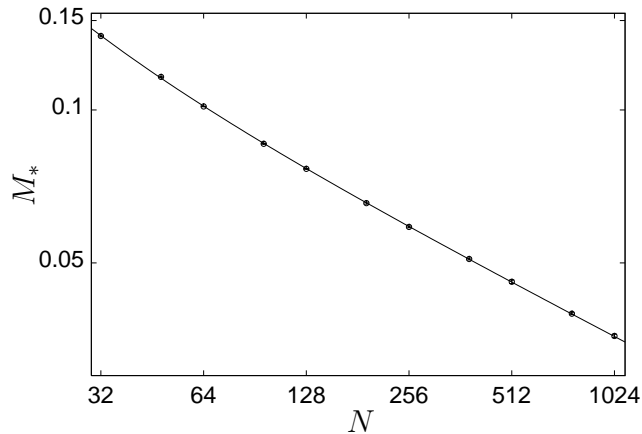}
\caption{\label{scM} The mass gap $M_*$ for different
system sizes $N$, and the curve representing
the best fit to the formula (\ref{scaling})~:
$1/d_H=0.348(4)$, $b=0.40(1)$ and $t=5.7(5)$.}
\end{center}
\end{figure}

The exponent $d_H$ is the fractal dimension
of the surface given. A typical linear extent 
of $L$ scales as $L=N^{1/d_H}$.
When the physical
mass is equal zero, $L$ sets the scale
for the correlation length. Its inverse gives
the minimal eigenvalue of the spectrum $M_*$.
For smaller systems one expects corrections to scaling.
We took it into account by introducing a phenomenological correction $t/N$
to the formula (\ref{scaling}). This correction significantly
improves quality of the fit for the studied range of $N$.
The best fit to the formula is
$1/d_H=0.348(4)$, $b=0.40(1)$ and $t=5.7(5)$. The corresponding
curve is plotted in fig.\ref{scM}. The curve
fits indeed very well to all the data points. The error bars of the
best fit parameters were estimated by jack-knife.

We compared the goodness of the best fits
to the formula (\ref{scaling}) and analogous
formulas in which the correction $t/N$
was substituted by $t/N^{1/2}$ and $t/N^{3/2}$.
We obtained $\chi^2/{\rm d.o.f.} =0.52$
for $t/N$ while in the other two cases
$1.66$ and $1.83$, respectively. Thus, among those
three correction types, the one $t/N$ 
is best in this range. We have also checked that the
fitted value $1/d_H=0.348(4)$ is stable against the
successive removal of the data points of the smallest volumes.

There are different theoretical predictions for the value
of the Hausdorff dimension $d_H=4$ \cite{ctbj,kkmw} and 
$d_H=3$ \cite{kn}, the latter of which was obtained for
a test fermion in the gravitational background coupled
to matter field with the central charge $c=1/2$ \cite{kn}.
The Hausdorff dimension measured in our MC simulations
$d_H=2.87(3)$ favors $d_H=3$.
One should, however, be aware that in 
measurements of the Hausdorff dimension there 
are large finite size effects,
as can be seen from the considerations 
of pure gravity \cite{ctbj}.

\begin{figure}
\begin{center}
\psfrag{xx}{$1/N$}
\psfrag{yy}{$K_*$}
\includegraphics[width=8cm]{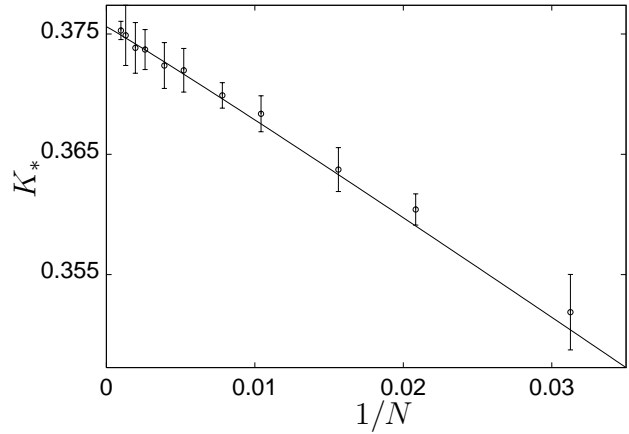}
\caption{\label{scK} The pseudo-critical parameter $K_*$
for different system sizes $N$, and the curve representing
the best fit to the scaling formula (\ref{scaling})~:
$K_\infty=0.3756(16)$, $\kappa=1.03(30)$
$a=-0.9(5)$, plotted as a function of $1/N$.}
\end{center}
\end{figure}

The best fit for the second formula in (\ref{scaling})
is given by $K_\infty=0.3756(16)$, $\kappa=1.03(30)$ and
$a=0.9(5)$ (see fig.\ref{scK}).
The limiting value $K_\infty$ is in
agreement with the theoretically calculated critical
value $K_{cr}$ (\ref{crit}). The scaling exponent
$\kappa$ is almost equal $1$ which would suggest
a kinematic scaling saying
that the average distance between eigenvalues
decreases as $1/N$. 

\section*{Conclusions and outlook}

We have investigated the properties of the Dirac-Wilson operator
on a random triangulation. In particular we have 
extracted from the spectrum of the operator
the values of the fractal dimension $d_H$ and
the critical value of the hopping parameter $K_{cr}$.
At this value of $K=K_{cr}$ the fermions become massless
and one can take a continuum limit corresponding to massless
fermions interacting with $2d$ gravity.
The value of $K_{cr}$ has been calculated analytically, however,
$d_H$ has not been unambiguously determined theoretically. 
In the neighborhood of $K_{cr}$, which lies deep in
the ferromagnatic phase we found spectral
distributions which are typical for random systems.

Apart from $K_{cr}$ there is another interesting value
of $K$, namely $K_0= 1/\sqrt{3}$ which lies on the
border between the ferromagnetic and antiferromagnetic
phases of the corresponding Ising model. 
We observed that the distributions of the lowest
eigenvalues, $\widehat{\rho}_j(x)$, becomes discrete when $K$ goes 
to $K_0$ from below.
This is an unexpected phenomenon for a random system.

For some values of the hopping parameter $K$ in the
'antiferromagnetic' phase,  eigenvalues of 
the Dirac-Wilson operator decouple from the
random structure of the matrix and depend only
on local geometrical properties of the triangulation. 
This, as we discussed, leads to the appearance
of discrete spectra at $K_0$.

There are many natural extensions of the studies presented
in this work. One should try to understand
properties of the spectrum of the Dirac operator
from the point of view of the random matrix theory \cite{m,vz,vw}.
This is a slightly
different type of randomness than the one provided by the coupling
to the vector gauge field which is usually discussed in the
context of QCD. However exactly this type of randomness may be
important in quantum gravity.

Next, one can investigate in more detail the relation of the quenched
approximation to the full model. 
The quenched model describes a test particle 
in pure gravity. From this exercise one could 
perhaps draw a general lesson
about the effect of quenching on the spectrum of
the Dirac operator. This can be important because
this type of approximation is frequently used in many
physical contexts, for example, in QCD. However, one usually
is not able to quantify the effects of quenching.

Furthermore, one can study effects of changing topology
by considering non-spherical 2d-triangulations.
As mentioned, this requires a careful treatment of various
spin structures which may be admitted
by a manifold. Contrary to higher dimensional case, where
the existence of spin structure is related to the
second Stiefel-Withney class \cite{top}, here the question of the
existence reduces to the orientability of the manifold.
Also the classification of spin structures is relatively simple
in the 2d case. The spin structures can be classified by
a set of signs defined on all classes of non-contractable loops.
The signs tell us whether boundary conditions for a
parallel transport of a spinor around those loops are periodic ($+1$)
or anti-periodic ($-1$).
For a manifold with genus $h$,
there are $2h$ different classes of non-contractable
loops and hence there are $2^{2h}$ different
spin structures.

Finally one should try to find a lattice implementation of the
Dirac operator for higher dimensional compact
simplicial manifolds. Many parts of the construction can be directly
generalized from the 2d case; actually almost all, except
the link sign degrees of freedom, $s_{ij}$ (\ref{u}),
which as it turns out are not sufficient in general case
for the connections ${\cal U}_{ij}$
to fulfill the consistency condition for all plaquettes (\ref{sp}).

\section*{Acknowledgments}
We thank J. Jurkiewicz, A. Krzywicki, E. Laermann,
D. E. Miller and J. Tabaczek for many discussions. This work was
supported in part by the EC IHP grant HPRN-CT-1999-00161 and
by the Polish Government Project (KBN)  2P03B 01917.
At the first stage of this work L.B. was supported by
a DAAD fellowship.

\section*{Appendix}
\begin{figure}
\begin{center}
\psfrag{e1}{X}
\psfrag{e2}{Y}
\includegraphics[width=6cm]{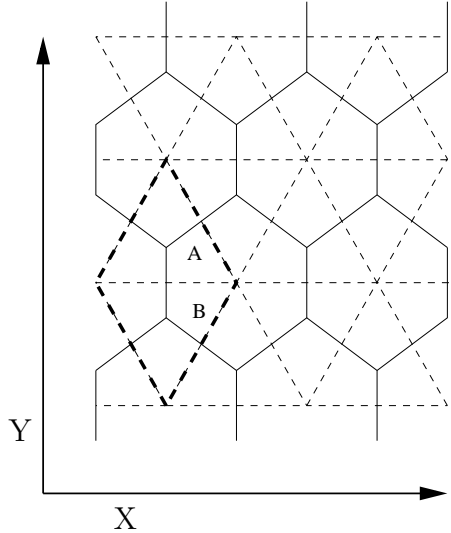}
\caption{\label{honey} Regular triangulation of the plane
and its dual lattice. Fermions live on the vertices of the
dual (honey-comb) lattice. The elementary cell
contains two distinct dual node positions denoted by
$A$ and $B$. }
\end{center}
\end{figure}

For a comparison in the appendix
we calculate spectrum
of the Dirac-Wilson operator on the regular
planar triangulation built of
equilateral triangles with fermion field
located in the centers of triangles.
If one connects the centers by links, they
form a dual lattice;
in this case it is a honey-comb lattice
(see fig.\ref{honey}).
It is convenient to divide the vertices of this lattice
into two classes $A$ and $B$
forming a check-board. The fundamental cell on the
triangulation contains one site of each.
One reconstructs the entire triangulation
translationally copying the fundamental
cell using multiples of two the vectors
$d_1=n_0+n_1$, and $d_2=n_0+n_2$ constructed from
the link vectors $n_0=(0,1)$,
$n_1=(\sqrt{3}/2,1/2)$, $n_2=(-\sqrt{3}/2,1/2)$.

\begin{figure}
\begin{center}
\psfrag{xx}{Re}
\psfrag{yy}{Im}
\includegraphics[width=8cm]{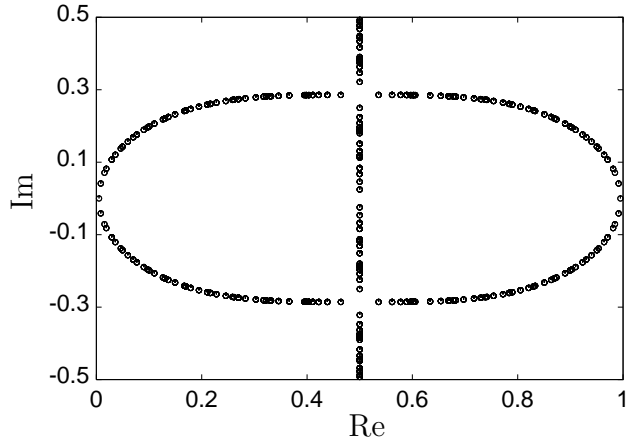}
\caption{\label{okulary} Eigenvalues $\lambda$
of the Dirac-Wilson operator on a regular triangulation
with $L=50$ and $K=0.33$.}
\end{center}
\end{figure}

Using translational symmetry of the lattice we can
now rewrite the action (\ref{sh}) in the following form
\begin{eqnarray}
S & = & -\frac{K}{2} \sum_{i} \sum_{d = 1}^2 \left[
\bar{\psi}_{i+d, A} (1 + n_d \cdot \gamma)
\psi_{i, B} +
\bar{\psi}_{i, B} (1 - n_d \cdot \gamma)
\psi_{i + d, A} \right] \nonumber \\
& & -\frac{K}{2} \sum_{i} \left[
\bar{\psi}_{i, A} (1 - n_0 \cdot \gamma)
\psi_{i, B} +
\bar{\psi}_{i, B} (1 + n_0 \cdot \gamma)
\psi_{i, A} \right] \nonumber \\
& & + \frac{1}{2} \sum_{i} \left[
\bar{\psi}_{i, A} \psi_{i, A} +
\bar{\psi}_{i, B} \psi_{i,B} \right] \, ,
\label{hc}
\end{eqnarray}
where the first index in $\psi_{i,A}$ is a double
index consisting of two integers $(i_1,i_2)$, which
give the position of the cell
$x = i_1 d_1 + i_2 d_2$, while the second label denotes
the position A or B within the cell. In the component
notation, the addition of $d_1$ to $i$ corresponds to
$(i_1,i_2) \rightarrow (i_1+1,i_2)$, and of $d_2$ to
$(i_1,i_2) \rightarrow (i_1,i_2+1)$. In the expression
(\ref{hc}) we have used a shorthand notation denoting
the sum over $d_1$ and $d_2$ by $d=1,2$.
We can now partially diagonalize the problem
using the Fourier transform of the index
$i=(i_1,i_2)$ to the momentum space $p=(p_1,p_2)$. This
leads us to a block-diagonal matrix consisting
of four by four blocks. Each block $D(p)$ corresponds to
one Fourier mode $\bar{\psi}_p D(p) \psi_p$. The four by four
matrix $D(p)$ is indexed by the spinor index of $\psi$
and of the position label A or B.
For each $p$, diagonalization of $D(p)$ yields four eigenvalues
\begin{equation}
\lambda_p = \frac{1}{2} \pm K \frac{\sqrt{3}}{2}
\sqrt{w \pm i\sqrt{4 - (w - 1)^2}} \, ,
\label{lambda}
\end{equation}
where
\begin{equation}
w = \cos (p_1) + \cos (p_2) + \cos (p_1 - p_2) \, .
\end{equation}
The distribution of eigenvalues (\ref{lambda})
on a finite lattice $L\times L$ with periodic
boundary condition in the $d_{1,2}$ directions
is shown in fig.\ref{okulary}. In this case the momenta
admit the values $p_{1,2} = 2\pi k_{1,2}/L$,
where $k_{1,2}=0,\dots, L-1$ and hence the operator
has $4L^2$ eigenvalues.

\begin{figure}
\begin{center}
\psfrag{xx}{$\widehat\lambda$}
\psfrag{yy}{$\widehat\rho$}
\includegraphics[width=8cm]{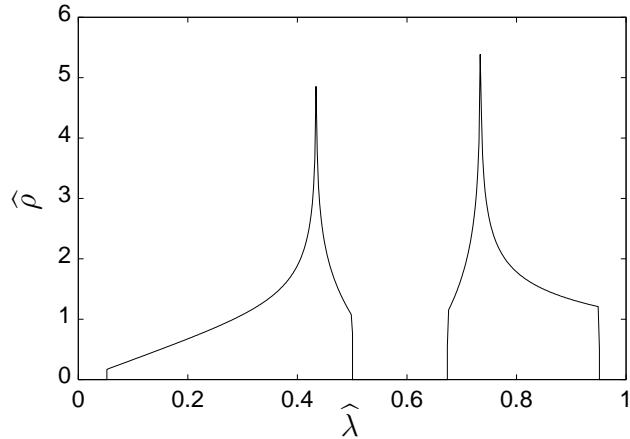}
\caption{\label{hhs} Histograms of
eigenvalues of the operator $\widehat{\cal D}$
for $L=3000$ and $K=0.3$. At the critical value $K=K_{cr}$ the spectrum goes
continuously to zero, while  for $K\ne K_{cr}$, like for example for $K=0.3$
presented in the figure, 
its low-$\lambda$ part is cut off at some positive $\lambda_{min}$.}
\end{center}
\end{figure}

Similarly, one can find eigenvalues
of the operator $\widehat{\cal D}$
\begin{equation}
\widehat{\lambda}_p =
\pm \frac{i}{\sqrt{2}}
\sqrt{\frac{1}{2}+\frac{K^2}{2}(w+6) \pm
\sqrt{(\frac{K^2}{2}(w-3)+2)^2+9K^2-4}} \, .
\label{hatlambda}
\end{equation}
Using this formula we can calculate the spectral density
\begin{equation}
\widehat{\rho}(x) = \lim_{L\rightarrow \infty}
\frac{1}{4L^2} \sum_\lambda \delta(x - i \widehat{\lambda}) \, .
\end{equation}
The spectrum terminates at a small positive value $K_{cr}$
(see fig.\ref{hhs}). It goes to zero only for $K=K_{cr}=\frac{1}{3}$.
In the large $L$-limit, the two peaks
in fig. \ref{hhs} develop a logarithmic singularity.

For the regular lattice the critical value
of the hopping parameter is given by the standard
equation $K_{cr} = 1/q$, where $q$ is
the number of links emerging from the vertex. In this case $q=3$.
For random lattice this condition is dressed by lattice
fluctuations. Although each vertex has coordination $q=3$,
the critical value of the hopping parameter is shifted from $1/3$
to the value given by equation (\ref{crit}).
For the regular lattice, the spectrum
has an eigenvalue equal exactly zero for the critical value of the
hopping parameter. This is not the case for random lattice,
where the smallest eigenvalue has a distribution whose center
of mass approaches zero only for large $N$.
On the regular lattice, the lowest part of the spectrum does
not move when $N$ goes to infinity, but becomes denser.
The average distance between the eigenvalues
scales like $N^{-1/d_H}$, with the canonical dimension
$d_H=2$ while on the random lattice, the position of
the lowest eigenvalue moves towards zero with
$N^{-1/d_H}$ with a dressed exponent $d_H=2.87(3)$
resulting from the fractal structure of fluctuating geometry.

\end{document}